\documentclass[transmag]{IEEEtran}
\usepackage{latexsym}
\usepackage{graphicx}
\usepackage{amsfonts,amssymb,amsmath}
\usepackage{hyperref}
\usepackage{cite}
\usepackage{algorithmic}
\usepackage{graphicx}
\usepackage{textcomp}
\usepackage{xcolor}
\usepackage{float}
\usepackage{amsthm}
\usepackage{graphicx}
\usepackage{epstopdf}
\usepackage{amsmath,bm}
\usepackage{amsfonts}
\usepackage{amssymb}
\usepackage{color}
\usepackage{multirow}
\usepackage{multicol}
\usepackage{soul,xcolor}
\usepackage{overpic}
\definecolor{blue}{rgb}{0,0,0}

\def\BibTeX{{\rm B\kern-.05em{\sc i\kern-.025em b}\kern-.08em T\kern-.1667em\lower.7ex\hbox{E}\kern-.125emX}}
\begin{document}
\newcommand*{\rom}[1]{\expandafter\@slowromancap\romannumeral #1@}
\title{Channel Estimation in Massive MIMO under Hardware Non-Linearities: Bayesian Methods versus Deep Learning}

\author{\"Ozlem Tugfe Demir,~\IEEEmembership{Member,~IEEE,}  Emil Bj\"ornson,~\IEEEmembership{Senior Member,~IEEE}
\thanks{This work was partially supported by ELLIIT and the Wallenberg AI, Autonomous Systems and Software Program (WASP) funded by the Knut and Alice Wallenberg Foundation. A part of this paper was presented in International Symposium on Wireless Communication Systems 2019 \cite{ozlem_iswcs}.}
\thanks{The authors are with the Department of Electrical Engineering
	(ISY), Linköping University, 581 83 Linköping, Sweden (e-mail: ozlem.tugfe.demir@liu.se, emil.bjornson@liu.se)}}

\IEEEtitleabstractindextext{\begin{abstract}
	This paper considers the joint impact of non-linear hardware impairments at the base station (BS) and user equipments (UEs) on the uplink performance of single-cell massive MIMO (multiple-input multiple-output) in practical Rician fading environments. First, Bussgang decomposition-based effective channels and distortion characteristics are analytically derived and the spectral efficiency (SE) achieved by several receivers are explored for third-order non-linearities. Next, two deep feedforward neural networks are designed and trained to estimate the effective channels and the distortion variance at each BS antenna, which are used in signal detection. We compare the performance of the proposed methods with state-of-the-art distortion-aware and -unaware Bayesian linear minimum mean-squared error (LMMSE) estimators. The proposed deep learning approach improves the estimation quality by exploiting impairment characteristics, while LMMSE methods treat distortion as noise. Using the data generated by the derived effective channels for general order of non-linearities at both the BS and UEs, it is shown that the deep learning-based estimator provides better estimates of the effective channels also for non-linearities more than order three.
\end{abstract}


}

\maketitle

\section{INTRODUCTION \label{introduction}}

\IEEEPARstart{M}{assive}  MIMO (multiple-input multiple-output) with a large number of antennas and fully digital transceivers at the base stations (BSs), is now a practical technology whose main concepts are adopted to 5G \cite{towards_massive_mimo}. Channel estimation using the uplink pilot sequences in both conventional and massive MIMO is a well-studied problem \cite{Kotecha2004a,Neumann2018,emil_book} in the case of ideal hardware at both the BS and user equipments (UEs). However, in practice, transceiver impairments, such as non-linearities in amplifiers, I/Q imbalance, and quantization errors are inevitable \cite{book_rf}. Some papers in the massive MIMO literature model the continuous hardware impairments using a stochastic additive model \cite{emil_nonideal,stochastic_model1,stochastic_model2,rician_additive}. However, behavioral models which utilize some deterministic functions are expected to model the continuous non-linear distortion better and are used in many different research areas \cite{behavioral_model1,christopher,pa,bussgang2,ozlem_iswcs,nonlinear1,nonlinear2,nonlinear3, nonlinear4, sina, peter1, peter2}.

The non-linear system behavior is often treated by utilizing the Bussgang decomposition to find an equivalent linear system with uncorrelated distortion \cite{emil_nonideal,bussgang1, bussgang2,peter1,peter2,sina,bussgang3}. One can then derive a distortion-aware Bayesian LMMSE estimator that utilizes the first- and second-order distortion statistics to estimate the channels, but in doing so the distortion is treated as independent colored noise, although it depends on the channel. Furthermore, we should note that deriving the minimum mean-squared error (MMSE) estimator is usually very hard in the case of non-linear hardware impairments. Hence, this brings the need to design new methods to beat the conventional Bayesian estimators by exploiting the structure of the impaired signal by hardware non-linearities and, particularly, that the distortion is dependent on the desired signal. 

There are several works which model and analyze the impact of hardware non-linearities on massive MIMO using behavioral modeling \cite{christopher, behavioral_model1,bussgang2,emil_hardware,sina,nonlinear1,nonlinear2,nonlinear3,nonlinear4}. Recently,\cite{emil_hardware} proposed several distortion-aware receivers for uplink signal detection in massive MIMO. To apply these receivers, it is necessary for the BS to know the effective channels of the UEs together with the received signal correlation matrix. This has motivated us to consider the estimation of the effective channels, taking into account the BS and UE non-linear distortion characteristics, instead of only the wireless channels. To the best of authors' knowledge, this paper is the first work which considers channel estimation under both BS and UE non-linear distortions by using quasi-memoryless polynomial modeling.

\subsection{Main Contributions}
The first novelty of this paper is the derivation of the effective channels and distortion correlation matrix for arbitrary symmetric finite-sized constellations in the uplink data transmission when the BS and UEs are subject to third-order quasi-memoryless polynomial distortion. Note that this model can represent both amplitude-to-amplitude modulation (AM/AM) and amplitude-to-phase modulation (AM/PM) distortions and  is used in accordance with previous literature \cite{peter1,peter2,christopher,pa,ericsson}. We generalize the spectral efficiency (SE) analysis in \cite{emil_hardware} by taking the non-linear distortion at the UEs into account.  

As a second contribution, we derive the distortion-aware LMMSE-based channel estimator analytically for Rician fading. Then, we utilize the derived analytical models to design  novel deep-learning-based estimators of the effective channels and distortion variances to implement several uplink receivers. We train the neural networks to exploit the full structure of the hardware impairments, instead of treating the distortion as independent noise as in previous work. We compare our novel solutions with both distortion-aware and unaware LMMSE estimators and show that the deep-learning-based alternatives significantly outperform them.

Thirdly, we generalize the hardware impairment model to higher-order quasi-memoryless polynomials and derive the analytical expressions for effective channels and train the deep learning network we propose for the effective channel estimation for any order non-linear distortions at both the BS and UEs.

Note that the prior conference version of this paper \cite{ozlem_iswcs} considers only channel estimation without UE non-linearities in Rayleigh fading and do not include the distortion variance estimator and the analytical results for the distortion-aware LMMSE-based channel estimator.

\emph{Reproducible research:} All the simulation results can be reproduced using the Python code and data files available at: https://github.com/emilbjornson/deep-learning-channel-estimation.

\section{System Model with BS and UE Hardware Impairments \label{system}}
We consider a single-cell massive MIMO system where a BS equipped with $M$ antennas serves $K$ single-antenna UEs simultaneously. In this paper, we focus on the uplink to mitigate the adverse effects of non-ideal BS and UE hardware on the system performance. A block-fading model is considered where the channels between each antenna of the BS and UEs can be represented by a constant complex-valued scalar that takes an independent realization in each time-frequency coherence block \cite{emil_book}. In each block, the channels are estimated via uplink training using pilot sequences. Then, the estimated channels are used for signal detection during uplink data transmission.  

In any arbitrary coherence block, the noise and BS-distortion-free signal ${\bf u}=[ u_1 \ ... \ u_M]^T \in \mathbb{C}^M$ at the input of the receive BS antennas in the data transmission phase is 
\begin{align}\label{eq:u}
{\bf u}=\sum_{k=1}^K{\bf g}_ks_k={\bf G}{\bf s},
\end{align}
where ${\bf G}=[{\bf g}_1 \ \ldots \ {\bf g}_K] \in \mathbb{C}^{M\times K}$ is the concatenated channel matrix where ${\bf g}_k=[g_{k1}\ \ldots \ g_{kM}]^T \in \mathbb{C}^M$ is the channel from the $k^{\textrm{th}}$ UE to the BS. $s_k \in  \mathbb{C}$ is the information-bearing signal of the $k^{\textrm{th}}$ UE and ${\bf s}=[s_1 \ \ldots \ s_K]^T\in \mathbb{C}^{K}$. It is assumed that all information signals are independent  and $\mathbb{E}\{|s_k|^2\}=p_k$, for $k=1,\ldots,K$, where $p_k$ is the transmission power of the $k^{\textrm{th}}$ UE. 

In this paper, we consider spatially uncorrelated Rician fading channels where each channel vector ${\bf g}_k$, for $k=1,\ldots,K$, follows the complex Gaussian distribution
\begin{align}\label{eq:gk}
{\bf g}_{k}\sim \mathcal{N}_{\mathbb{C}}(\bar{{\bf g}}_k,\beta_k{\bf I}_M), 
\end{align}
where the mean vector $\bar{{\bf g}}_k \in \mathbb{C}^M$ models the LOS component of the $k^\textrm{th}$ UE. The zero-mean parts of the channels are circularly symmetric Gaussian random variables and they model small-scale fading which is assumed to be uncorrelated among the BS antennas. $\beta_k$ is the large-scale fading coefficient and it describes the long-term channel effects such as pathloss and shadowing. The channel statistics $\{\bar{{\bf g}}_k\}$ and $\{\beta_k\}$ are assumed to be known at the BS in accordance with the massive MIMO literature \cite{emil_book}, but practical estimation methods are described in \cite{towards_massive_mimo}.

Note that in the first part of this paper, we will explore the effect of  non-linearities in the UEs' and BS's radio frequency (RF) hardware and derive the effective channels for a given realization of the channels. Hence, we will not use the statistical distributions of the channels. However, in the second part, we will derive the distortion-aware LMMSE estimator by considering the first- and second-order statistics of the channels and train a deep neural network to perform channel estimation with samples according to the distribution in \eqref{eq:gk}. 
We will now investigate the effect that non-ideal hardware has on ${\bf u}$.
\subsection{Quasi-Memoryless Polynomial Modeling of BS Hardware Impairments}
Unlike most of the previous works which utilize stochastic additive or multiplicative models \cite{stochastic_model1, stochastic_model2,emil_nonideal}, we use in this paper a  more refined deterministic behavioral model for the hardware non-linearities for a more accurate modeling of the main sources of hardware impairments \cite{behavioral_model1}. {\color{blue} One of the major advantages of the deterministic behavioral models is their ability to model the physical effect of hardware impairments on the baseband signals for various implementations using a small number of parameters. In fact, the modeling does not depend on a specific RF front-end, but the same non-linear models can be used with a sufficiently good accuracy for measuring the performance \cite{book_rf}. In this paper,} the non-ideal BS receiver hardware is modeled as a behavioral non-linear quasi-memoryless function where both the amplitude and phase of the received signal are distorted. {\color{blue} Considering only the memoryless non-linearities is a meaningful assumption for moderate bandwidths such as 20\,MHz \cite{book_rf}. Furthermore, it is analytically manageable to derive the moments of the distorted signals using the considered quasi-memoryless functions.} In the first part of the paper, we will use the following third-order polynomial model for this kind of distortion in the complex baseband \cite{book_rf}:
\begin{align} \label{eq:z_m}
z_m=\tilde{a}_{0m}u_m+\tilde{a}_{1m}|u_m|^{2}u_m, \ \ m=1,\ldots,M,
\end{align}
where $z_m$ is the noise-free distorted signal at the $m^\textrm{th}$ BS antenna and $\{\tilde{a}_{0m}, \tilde{a}_{1m}\}$ are complex scalar coefficients, which means that both AM/AM and AM/PM distortions are considered \cite{book_rf}. The 
model in \eqref{eq:z_m} jointly describes the non-linearities in amplifiers, local oscillators, mixers, and other hardware components. {\color{blue} Note that for the in-band distortion, the quasi-memoryless polynomials only have the odd order terms since the even order terms appear out of band \cite{book_rf} and third-order terms capture the main source for the RF amplifiers \cite{peter1,peter2}.} We further assume that long-term automatic gain control is utilized, thus $\tilde{a}_{lm}$ can be represented by
\begin{align} \label{eq:tilde-alm} \tilde{a}_{lm}&=\frac{a_{lm}}{{\big(b^{\text{BS}}_{\text{off}}\mathbb{E}\{|u_m|^2\}\big)^l}}\nonumber \\
&=\frac{a_{lm}}{\big(b^{\text{BS}}_{\text{off}}\sum_{k=1}^K(|\bar{g}_{km}|^2+\beta_k)p_k\big)^l}, \ \ l=0,1,
\end{align}
where $\{a_{lm}\}$ are normalized reference polynomial coefficients when the input signal to the receiver has a magnitude between zero and one \cite{ericsson}. In practical communication systems, a backoff is applied in the low-noise amplifier (LNA) to prevent clipping due to the nonlinear components. Here, $b^{\text{BS}}_{\text{off}}$ is the backoff parameter at the BS. Using \eqref{eq:z_m}, the digital baseband signal ${\bf y}=[y_1 \ \ldots \ y_M]^T \in \mathbb{C}^{M}$ at the BS is given by
\begin{align}\label{eq:digitial-baseband}
{\bf y}={\bf z}+{\bf n},
\end{align} 
where  ${\bf z}=[ z_1 \ \ldots \ z_M]^T \in \mathbb{C}^{M}$ is the hardware-distorted signal from \eqref{eq:z_m} and ${\bf n} \sim \mathcal{N}_{\mathbb{C}}({\bf 0}_M,\sigma^2{\bf I}_M)$ is uncorrelated noise. In practice, the initial noise entering into the BS hardware is also affected by the nonlinear distortion, however the resultant noise is still uncorrelated with ${\bf u}$ \cite{emil_hardware}.

Now, we will analyze the UE hardware impairment effect on the information-bearing signals.
\subsection{Modeling of UE Hardware Impairments}
Some works in the literature assume perfect UE hardware \cite{christopher} when analyzing the BS distortion. However, as shown in \cite{emil_book,emil_nonideal,emil_hardware}, UE hardware impairments can be the performance-limiting factor since it is not averaged out over the BS antennas. Most of the works in the literature assume stochastic additive model for the UE hardware distortion. Note that in \cite{emil_hardware}, the effect of third-order non-linearities due to the BS hardware is analyzed where a stochastic additive distortion is assumed at each UE. In this model, the UE hardware distortion is independent of the uplink data signals. A more realistic approach is to use a deterministic third-order behavioral model also at the UEs to take into account the possible non-linearities and symbol-dependent distortion. The first novelty of this paper is to  study the effect of non-linearities in a symbol-sampled system by adopting a behavioral model at both the BS and UEs. Even if some predistortion is applied, we can model the residual non-linearities at the UE side by using a third-order quasi-memoryless polynomial model  
\begin{align}\label{eq:sk}
s_k=\sqrt{\eta_k}\big(\tilde{b}_{0}\varsigma_k+\tilde{b}_{1}|\varsigma_k|^2\varsigma_k\big),
\end{align}
{\color{blue} by following the same reasoning for modeling the BS hardware impairments. In fact,
	the third order intercept point, which can be related to the coefficient of the third-order term in \eqref{eq:sk}, is a common quality measure for the distortion in RF amplifiers \cite{peter1}. In \eqref{eq:sk}}, $\varsigma_k$ is the actual desired signal to be transmitted from the $k^{th}$ UE with zero mean and $\mathbb{E}\{|\varsigma_k|^2\}=1$. The complex scalar coefficients $\{\tilde{b}_{0},\tilde{b}_{1}\}$ are given by
\begin{align}\label{eq:bk}
\tilde{b}_{l}=\frac{b_{l}}{{\big(b^{\text{UE}}_{\text{off}}\mathbb{E}\{|\varsigma_k|^2\}\big)^l}}=\frac{b_{l}}{\big(b^{\text{UE}}_{\text{off}}\big)^l}, \ \ l=0,1,
\end{align}
where $\{b_0,b_1\}$ are the normalized reference polynomial coefficients when the input signal to the transmitter has a magnitude between zero and one{\color{blue} \cite{ericsson}}. $\sqrt{\eta_k}$ is the scaling factor such that $\mathbb{E}\{|s_k|^2\}=p_k$ under the assumption that variable-gain power amplifiers are used at the UEs. Note that the operating point of the power amplifier of UEs is adjusted with an input backoff $b_{\text{off}}^{\text{UE}}$ which can be the same for all the UEs since we use a normalized distortion model. Note that hardware distortion characteristics  is assumed to be the same for all the UEs for analytical tractability.

For notational convenience, we now define the distorted transmit signal without power scaling as $\upsilon_k\triangleq\tilde{b}_0\varsigma_k+\tilde{b}_1|\varsigma_k|^2\varsigma_k$, for $k=1,\ldots,K$. We will use this definition throughout the paper. We assume the same modulation for all the UEs{\color{blue}, which is particularly useful in massive MIMO systems that aim to provide uniformly good service to all the UEs. Under the same modulation assumption,} the symbols $\upsilon_k$, for $k=1,\ldots,K$, are independent and identically distributed (i.i.d.). This property will allow us to obtain analytically tractable results,  However, the proposed deep learning based channel estimation can be used for offline deep learning training for UEs possibly having different modulations. 

If we further define
\begin{align} \label{eq:moments1}
\zeta_l\triangleq\mathbb{E}\{|\varsigma_k|^{l}\}, \ \ l=2,4,6,\ldots,
\end{align}
we can easily find the even order moments of zero-mean i.i.d. variables $\upsilon_k=\tilde{b}_0\varsigma_k+\tilde{b}_1|\varsigma_k|^2\varsigma_k$ using \eqref{eq:moments1}. The even order moments of $\upsilon_k$ are defined as follows:
\begin{align} \label{eq:moments2}
&\chi_l\triangleq\mathbb{E}\{|\upsilon_k|^l\}, \ \ \ l=2,4,6, \ldots.
\end{align}
The power scaling parameter $\eta_k$ in \eqref{eq:sk} can be found by evaluating the average power of $s_k$ in \eqref{eq:sk} and equating it to $p_k$ as follows:
\begin{align}\label{eq:symbol-power}
&\mathbb{E}\{|s_k|^2\}=\eta_k\chi_2=p_k \Rightarrow \eta_k=\frac{p_k}{\chi_2}.
\end{align}

In \cite{emil_hardware}, it is assumed that the actual desired information signals $\{\varsigma_k\}$ are  complex Gaussian in order to maximize differential entropy when evaluating SE. Different from \cite{emil_hardware}, we will also consider symmetric finite-sized constellation for the information signals and design distortion-aware receivers for signal detection. This is the second novelty of this paper. In the next part, we will derive the effective channels for a general class of information signals.

\section{Effective Channels for Third-Order Non-linearities \label{effective}}

We consider one fixed channel realization ${\bf G}$ in an arbitrary coherence block and let $\mathbb{E}_{|{\bf G}}\{.\}$ denote the conditional expectation given ${\bf G}$. Following the Bussgang decomposition approach \cite{bussgang1, bussgang2, bussgang3, emil_hardware}, the digital baseband signal in \eqref{eq:digitial-baseband} can be written as a summation of the LMMSE estimate of ${\bf y}$ given $\bm{\varsigma}=[\varsigma_1 \ ... \ \varsigma_K]^T\in \mathbb{C}^K$ plus the additive distortion term as follows:
\begin{align}\label{eq:Bussgang}
{\bf y}={\bf C}_{y\varsigma}{\bf C}_{\varsigma\varsigma}^{-1}\bm{\varsigma}+\bm{\mu},
\end{align}
where  ${\bf C}_{y\varsigma} \in \mathbb{C}^{M \times K}$ and ${\bf C}_{\varsigma\varsigma} \in \mathbb{C}^{K \times K}$ are defined as ${\bf C}_{y\varsigma}=\mathbb{E}_{|{\bf G}}\{{\bf y}\bm{\varsigma}^H\}$ and ${\bf C}_{\varsigma\varsigma}=\mathbb{E}_{|{\bf G}}\{\bm{\varsigma}\bm{\varsigma}^H\}=\mathbb{E}\{\bm{\varsigma}\bm{\varsigma}^H\}$. Note that $\bm{\mu}={\bf y}-{\bf C}_{y\varsigma}{\bf C}_{\varsigma\varsigma}^{-1}\bm{\varsigma}$ and it is uncorrelated with $\bm{\varsigma}$ by construction. ${\bf C}_{\varsigma\varsigma}$ is by assumption given by ${\bf C}_{\varsigma\varsigma}={\bf I}_K$. We call $ {\bf C}_{y\varsigma}$ the effective channel since the signal term in \eqref{eq:Bussgang} is ${\bf C}_{y\varsigma}{\bf C}_{\varsigma\varsigma}^{-1}\bm{\varsigma} =  {\bf C}_{y\varsigma} \bm{\varsigma}$, thus the system effectively behaves as a non-distorted system with channel matrix ${\bf C}_{y\varsigma}$ and additive noise $\bm{\mu}$. Note that the effective channel ${\bf C}_{y\varsigma}$ is a non-linear function of the physical channel matrix ${\bf G}$ and symbol constellation. The $(m,k)$th element of ${\bf C}_{y\varsigma}$, i.e., $[{\bf C}_{y\varsigma}]_{mk}$ is the effective channel between the $k^{\textrm{th}}$ UE and the $m^{\textrm{th}}$ BS antenna, and it is given by
\begin{align}\label{eq:eff1}
[{\bf C}_{y\varsigma}]_{mk}&=\mathbb{E}_{|{\bf G}}\{y_m\varsigma_k^{*}\}=\mathbb{E}_{|{\bf G}}\{z_m\varsigma_k^{*}\}\nonumber \\
&=\tilde{a}_{0m}\mathbb{E}_{|{\bf G}}\{u_m\varsigma_k^{*}\}+\tilde{a}_{1m}\mathbb{E}_{|{\bf G}}\{|u_m|^2u_m\varsigma_k^{*}\}.
\end{align}
Let us find the expectations in the last term in the sequel. The first one is given by
\begin{align} \label{eq:expec1}
\mathbb{E}_{|{\bf G}}\{u_m\varsigma_k^{*}\}&=\sum_{l=1}^Kg_{lm} \mathbb{E}\{s_l\varsigma_k^{*}\}\nonumber \\&\stackrel{(a)}{=}g_{km}\mathbb{E}\{s_k\varsigma_k^{*}\}\stackrel{(b)}{=}g_{km}\sqrt{\eta_k}\big(\tilde{b}_{0}+\zeta_4\tilde{b}_{1}\big),
\end{align}
where we used the independence of the zero-mean data signals of different UEs in (a) and symbol moments defined in \eqref{eq:moments1} in (b), respectively. The second expectation in \eqref{eq:eff1} is given by
\begin{align} \label{eq:expec2}
&\mathbb{E}_{|{\bf G}}\{|u_m|^2u_m\varsigma_k^{*}\}\nonumber \\
&\hspace{0.5cm}=\mathbb{E}_{|{\bf G}}\Bigg\{\sum_{l_1=1}^Kg_{l_1m} s_{l_1}\sum_{l_2=1}^Kg_{l_2m}^{*}s_{l_2}^{*}\sum_{l_3=1}^Kg_{l_3m} s_{l_3}\varsigma_k^{*}\Bigg\} \nonumber \\
&\hspace{0.5cm}=\sum_{l_1=1}^Kg_{l_1m}\sum_{l_2=1}^Kg_{l_2m}^{*}\sum_{l_3=1}^Kg_{l_3m} \mathbb{E}\{s_{l_1}s_{l_2}^{*}s_{l_3}\varsigma_k^{*}\}.
\end{align}
We will evaluate the symbol moments $\mathbb{E}\{s_{l_1}s_{l_2}^{*}s_{l_3}\varsigma_k^{*}\}$ for Gaussian and finite-sized constellations. For ease of notation, let us define 
\begin{align}\label{eq:tildeg}
\tilde{g}_{km}\triangleq g_{km}\sqrt{\eta_k},
\end{align} 
which represents the channel gain with power control. Now, using \eqref{eq:expec1}, \eqref{eq:expec2}, and \eqref{eq:tildeg}, $[{\bf C}_{y\varsigma}]_{mk}$ can be expressed as
\begin{align}\label{eq:eff2}
[{\bf C}_{y\varsigma}]_{mk}&=\tilde{a}_{0m}\tilde{g}_{km}\big(\tilde{b}_{0}+\zeta_4\tilde{b}_{1}\big)\nonumber\\
&+\tilde{a}_{1m}\sum_{l_1=1}^K\tilde{g}_{l_1m}\sum_{l_2=1}^K\tilde{g}_{l_2m}^{*}\sum_{l_3=1}^K\tilde{g}_{l_3m} \mathbb{E}\{\upsilon_{l_1}\upsilon_{l_2}^{*}\upsilon_{l_3}\varsigma_k^{*}\}.
\end{align}
This expression holds for data signals that are either Gaussian or belong to the finite-sized constellation. We assume standard finite-sized constellations that satisfy the $90^{\circ}$ circular shift symmetry. This implies that if $\varsigma$ is a point in the constellation, then $\varsigma e^{j\frac{\pi}{2}s}$ for $s=1,2,3$ is also a constellation point. This kind of symmetry exists in most practically used constellations: PSK of dimension divisible by four, square QAM, circular QAM, etc. For these constellations, it is easy to prove that for any $l_1,l_2\in \mathbb{Z}^{+}$,  $\mathbb{E}\{\varsigma_k^{l_1}{\varsigma_k^{*}}^{l_2}\}=0$ if $l_1-l_2\neq 4i$ for any $i \in \mathbb{Z}$ under the equal symbol probability assumption. This property is also satisfied by circularly symmetric Gaussian data signals. Under the shift symmetry, it can easily be shown that for any $l_1,l_2\in \mathbb{Z}^{+}$, distorted symbols satisfy  $\mathbb{E}\{\upsilon_k^{l_1}{\upsilon_k^{*}}^{l_2}\}=0$ if $l_1-l_2\neq 4i$ for any $i \in \mathbb{Z}$. In this case, $\mathbb{E}\{\upsilon_{l_1}\upsilon_{l_2}^{*}\upsilon_{l_3}\varsigma_k^{*}\}$ in \eqref{eq:eff2} is given by the following lemma which is valid for Gaussian and symmetric finite-sized constellation data signals.

\emph{Lemma 1:} Suppose that for any $l_1,l_2\in \mathbb{Z}^{+}$, the i.i.d. random variables $\{\varsigma_k\}$ satisfy $\mathbb{E}\{\varsigma_k^{l_1}{\varsigma_k^{*}}^{l_2}\}=0$ if $l_1-l_2\neq 4i$ for any $i \in \mathbb{Z}$. Then, the moments $\mathbb{E}\{\upsilon_{l_1}\upsilon_{l_2}^{*}\upsilon_{l_3}\varsigma_k^{*}\}$ where $\upsilon_k=\tilde{b}_0\varsigma_k+\tilde{b}_1|\varsigma_k|^2\varsigma_k$, are given in \eqref{eq:symbol-gaussian} at the top of the following page,
\begin{figure*}
\begin{align}\label{eq:symbol-gaussian}
& \mathbb{E}\{\upsilon_{l_1}\upsilon_{l_2}^{*}\upsilon_{l_3}\varsigma_k^{*}\}=\begin{cases}
\zeta_{10}B_{1,1,1}+2\zeta_8B_{1,1,0}+\zeta_8B_{1,0,1}+2\zeta_6B_{0,0,1}+\zeta_6B_{0,1,0}+\zeta_4B_{0,0,0}, & \text{if } l_1=l_2=l_3=k, \\
(\tilde{b}_{0}+\zeta_4\tilde{b}_{1})(\zeta_6B_{1,1}+\zeta_4B_{1,0}+\zeta_4B_{0,1}+B_{0,0}), & \text{if } l_1=k\neq l_2=l_3, \\
(\tilde{b}_{0}+\zeta_4\tilde{b}_{1})(\zeta_6B_{1,1}+\zeta_4B_{1,0}+\zeta_4B_{0,1}+B_{0,0}), & \text{if } l_3=k\neq l_2=l_1, \\
0, & \text{otherwise}.
\end{cases}
\end{align}
\hrulefill
\end{figure*}
where 
\begin{align} 
&B_{r_1,r_2}\triangleq\tilde{b}_{r_1}\tilde{b}_{r_2}^{*}, \label{eq:B1} \\
&B_{r_1,r_2,r_3}\triangleq\tilde{b}_{r_1}\tilde{b}_{r_2}^{*}\tilde{b}_{r_3}. \label{eq:B2}
\end{align}
\begin{IEEEproof} Please see Appendix A.
\end{IEEEproof}
Using \eqref{eq:eff2} and Lemma 1 for symbol constellations that have the $90^{\circ}$ circular shift symmetry, the elements of the effective channel in \eqref{eq:eff2} for constellation with symbol moments $\zeta_l=\mathbb{E}\{|\varsigma_k|^l\}$, $l=2,4,6,\ldots$ are given by
\begin{align}\label{eq:eff-finite}
&[{\bf C}_{y\varsigma}]_{mk}=\tilde{a}_{0m}\tilde{g}_{km}\left(\tilde{b}_{0}+\zeta_4\tilde{b}_{1}\right)\nonumber\\&\hspace{0.5cm}+\tilde{a}_{1m}|\tilde{g}_{km}|^2\tilde{g}_{km}\Big(\zeta_{10}B_{1,1,1}+2\zeta_8B_{1,1,0}+\zeta_8B_{1,0,1}\nonumber\\&\hspace{0.5cm}+2\zeta_6B_{0,0,1}+\zeta_6B_{0,1,0}+\zeta_4B_{0,0,0}\Big)\nonumber \\
&\hspace{0.5cm}+2\tilde{a}_{1m}\tilde{g}_{km}\left(\tilde{b}_{0}+\zeta_4\tilde{b}_{1}\right)\times \nonumber \\
&\hspace{0.5cm}\left(\zeta_6B_{1,1}+\zeta_4B_{1,0}+\zeta_4B_{0,1}+B_{0,0}\right)\sum_{l=1, l\neq k}^K|\tilde{g}_{lm}|^2.
\end{align}

{\bf Remark:} The derived effective channels are valid for any channel model and depend on the instantaneous physical channels. Hence, they can be used for any channel model.

We have expressed the received signal at the BS in the form ${\bf y}={\bf C}_{y\varsigma}\bm{\varsigma}+\bm{\mu}$ and derived the elements of effective channel matrix ${\bf C}_{y\varsigma}$. In the following sections, we will analyze the SE of distortion-aware receivers using the derived effective channels, and then use \eqref{eq:eff-finite} for channel estimation.
\section{Spectral Efficiency \label{se}}

In this section, we quantify the performance of several distortion-aware receivers under a perfect channel state information (CSI) assumption while we consider classical and deep learning-based channel estimation schemes later. It is claimed in \cite{emil_hardware} that distortion correlation between BS antennas has negligible impact on the uplink {\color{blue} SE if the number of users is sufficiently large ($K>5$) and their SNR variations are relatively small. As an extension,} we will {\color{blue}quantify the gap between two linear receivers which either take into account the distortion correlation between different BS antennas or not} by analyzing third-order quasi-memoryless polynomial distortion at both BS and UEs.

Since we want to quantify the SE, we assume the uplink data signals are circularly symmetric Gaussian which maximizes the differential entropy, i.e., $\bm{\varsigma}\sim\mathcal{N}_{\mathbb{C}}({\bf 0}_K,{\bf I}_K)$. 

The elements of the effective channel matrix can easily be found by evaluating \eqref{eq:eff-finite} for Gaussian signals, i.e., $\zeta_l=(l/2)!$, for $l=2,4,\ldots$.

The distortion correlation matrix is given by
\begin{align}\label{eq:dist-corr}
{\bf C}_{\mu\mu}=\mathbb{E}_{|{\bf G}}\{\bm{\mu}\bm{\mu}^H\}={\bf C}_{zz}+\sigma^2{\bf I}_M-{\bf C}_{y\varsigma}{\bf C}_{y\varsigma}^H,
\end{align}
where ${\bf C}_{zz}=\mathbb{E}_{|{\bf G}}\{{\bf z}{\bf z}^H\}$ and we use the uncorrelatedness of $\bm{\varsigma}$ and $\bm{\mu}$. 

Before deriving the elements of ${\bf C}_{zz}$, we prove another lemma which is valid for Gaussian or symmetric finite-sized constellation data signals for ease of reference. 

\emph{Lemma 2:} Let ${\bf A}\in \mathbb{C}^{K \times K}$ and ${\bf B}\in \mathbb{C}^{K\times K}$ denote two deterministic matrices. For any $l_1,l_2\in \mathbb{Z}^{+}$ and $K$ zero-mean i.i.d. random variables $\{\upsilon_k\}$ such that $\mathbb{E}\{\upsilon_k^{l_1}{\upsilon_k^{*}}^{l_2}\}=0$ if $l_1-l_2\neq 4i$ for any $i \in \mathbb{Z}$, the following holds: 
\begin{align}
& \text{1) } \mathbb{E}\{\bm{\upsilon}\bm{\upsilon}^H{\bf A}\bm{\upsilon}\bm{\upsilon}^H\}\nonumber \\
&=\chi_2^2{\bf A}+\chi_2^2\text{tr}({\bf A}){\bf I}_K+(\chi_4-2\chi_2^2)\text{diag}({\bf A})\label{eq:lemma2-2}, \\
& \text{2) } \mathbb{E}\{\bm{\upsilon}\bm{\upsilon}^H{\bf A}\bm{\upsilon}\bm{\upsilon}^H{\bf  B}\bm{\upsilon}\bm{\upsilon}^H\}\nonumber \\
&=\chi_2^3\bigg({\bf A}{\bf B}+{\bf B}{\bf A}+\text{tr}({\bf A}){\bf B}+\text{tr}({\bf B}){\bf A}\nonumber \\
&\hspace{0.6cm}+\text{tr}({\bf A})\text{tr}({\bf B}){\bf I}_K+\text{tr}({\bf A}{\bf B}){\bf I}_K\bigg) \nonumber \\
&\hspace{0.6cm}+(\chi_4\chi_2-2\chi_2^3)\bigg(\text{diag}({\bf A}){\bf B}+\text{diag}({\bf B}){\bf A}+{\bf A}\text{diag}({\bf B})\nonumber\\
&\hspace{0.6cm}+{\bf B}\text{diag}({\bf A})+\text{diag}({\bf A}{\bf B}+{\bf B}{\bf A})+\text{tr}({\bf A})\text{diag}({\bf B})\nonumber \\ 
&\hspace{0.6cm}+\text{tr}({\bf B})\text{diag}({\bf A})+\text{tr}\big(\text{diag}({\bf A})\text{diag}({\bf B})\big){\bf I}_K\bigg) \nonumber \\
&\hspace{0.6cm}+(\chi_6-9\chi_4\chi_2+12\chi_2^3)\text{diag}({\bf A})\text{diag}({\bf B})\label{eq:lemma2-4},
\end{align}     
where $\bm{\upsilon}=[ \ \upsilon_1 \ \ldots \ \upsilon_K \ ]^T\in \mathbb{C}^K$ and the moments of $\upsilon_k$ are $\{\chi_l\}$ defined in \eqref{eq:moments2}.

\begin{IEEEproof} Please see Appendix B for the proof.
\end{IEEEproof}
The $(m,n)$th element of ${\bf C}_{zz}$  can be expressed as follows:
\begin{align}\label{eq:Czz}
&[{\bf C}_{zz}]_{mn}=\mathbb{E}_{|{\bf G}}\{z_mz_n^{*}\}\nonumber\\
&=\mathbb{E}_{|{\bf G}}\{(\tilde{a}_{0m}u_m+\tilde{a}_{1m}|u_m|^2u_m)(\tilde{a}_{0n}u_n+\tilde{a}_{1n}|u_n|^2u_n)^{*}\}\nonumber\\
&=\mathbb{E}_{|{\bf G}}\bigg\{\bigg(\tilde{a}_{0m}(\widetilde{\bf g}_m^{*})^H\bm{\upsilon}+\tilde{a}_{1m}|(\widetilde{\bf g}_m^{*})^H\bm{\upsilon}|^2(\widetilde{\bf g}_m^{*})^H\bm{\upsilon}\bigg)\times\nonumber\\
&\hspace{0.5cm}\bigg(\tilde{a}_{0n}(\widetilde{\bf g}_n^{*})^H\bm{\upsilon}+\tilde{a}_{1n}|(\widetilde{\bf g}_n^{*})^H\bm{\upsilon}|^2(\widetilde{\bf g}_n^{*})^H\bm{\upsilon}\bigg)^{*}\bigg\} \nonumber \\
&=\tilde{a}_{0m}\tilde{a}_{0n}^{*}\mathbb{E}_{|{\bf G}}\big\{(\widetilde{\bf g}_m^{*})^H\bm{\upsilon}\bm{\upsilon}^H\widetilde{\bf g}_n^{*}\big\}\nonumber\\
&+\tilde{a}_{1m}\tilde{a}_{0n}^{*}\mathbb{E}_{|{\bf G}}\big\{(\widetilde{\bf g}_m^{*})^H\bm{\upsilon}\bm{\upsilon}^H\widetilde{\bf g}_m^{*}(\widetilde{\bf g}_m^{*})^H\bm{\upsilon}\bm{\upsilon}^H\widetilde{\bf g}_n^{*}\big\} \nonumber \\
&+\tilde{a}_{0m}\tilde{a}_{1n}^{*}\mathbb{E}_{|{\bf G}}\big\{(\widetilde{\bf g}_m^{*})^H\bm{\upsilon}\bm{\upsilon}^H\widetilde{\bf g}_n^{*}(\widetilde{\bf g}_n^{*})^H\bm{\upsilon}\bm{\upsilon}^H\widetilde{\bf g}_n^{*}\big\}\nonumber \\
&+\tilde{a}_{1m}\tilde{a}_{1n}^{*}\mathbb{E}_{|{\bf G}}\big\{(\widetilde{\bf g}_m^{*})^H\bm{\upsilon}\times\nonumber \\
&\hspace{0.5cm}\bm{\upsilon}^H\widetilde{\bf g}_m^{*}(\widetilde{\bf g}_m^{*})^H\bm{\upsilon}\bm{\upsilon}^H\widetilde{\bf g}_n^{*}(\widetilde{\bf g}_n^{*})^H\bm{\upsilon}\bm{\upsilon}^H\widetilde{\bf g}_n^{*}\big\},
\end{align}
where $\widetilde{\bf g}_m\triangleq [ \ \tilde{g}_{1m} \ \ldots \ \tilde{g}_{Km} \ ]^T\in \mathbb{C}^K$, for $m=1,\ldots,M$. Note that the four terms in the last part of \eqref{eq:Czz} can be calculated using Lemma 2. 

In order to compute the SE using the above results, consider the combining vector ${\bf v}_k \in \mathbb{C}^M$ to be applied to the received signal ${\bf y}$ in \eqref{eq:digitial-baseband} for the $k^\textrm{th}$ user data signal detection. In this case, the instantaneous signal-to-interference-plus-noise ratio (SINR) for the $k^\textrm{th}$ user is given by
\begin{align}\label{eq:SINRk}
\text{SINR}_k=\frac{{\bf v}_k^H[{\bf C}_{y\varsigma}]_k[{\bf C}_{y\varsigma}]_k^H{\bf v}_k}{{\bf v}_k^H({\bf C}_{\mu\mu}+\sum_{i\neq k}[{\bf C}_{y\varsigma}]_i[{\bf C}_{y\varsigma}]_i^H){\bf v}_k},
\end{align} 
where $[{\bf C}_{y\varsigma}]_k$ denotes the $k^\textrm{th}$ column of the effective channel matrix ${\bf C}_{y\varsigma}$. Using \eqref{eq:SINRk}, the ergodic achievable SE $\mathbb{E}_{\bf G}\{\mathcal{I}(\varsigma_k;{\bf v}_k^H{\bf y})\}$ is lower bounded as \cite{emil_hardware}
\begin{align} \label{eq:SE}
\mathbb{E}_{\bf G}\{\mathcal{I}(\varsigma_k;{\bf v}_k^H{\bf y})\}\geq \mathbb{E}_{\bf G}\{\log_2(1+\text{SINR}_k)\},
\end{align}
where  $\mathbb{E}_{\bf G}\{.\}$ denotes the expectation with respect to physical channel matrix ${\bf G}$. In \cite{emil_hardware}, the distortion-aware MMSE (DA-MMSE) receiver is found by maximizing $\text{SINR}_k$ in \eqref{eq:SINRk} as follows:
\begin{align} \label{eq:da-mmse}
{\bf v}_k^{\text{DA-MMSE}}&=\big({\bf C}_{\mu\mu}+\sum_{i\neq k}[{\bf C}_{y\varsigma}]_i[{\bf C}_{y\varsigma}]_i^H\big)^{-1}[{\bf C}_{y\varsigma}]_k\nonumber\\
&=\big({\bf C}_{zz}+\sigma^2{\bf I}_M-[{\bf C}_{y\varsigma}]_k[{\bf C}_{y\varsigma}]_k^H\big)^{-1}[{\bf C}_{y\varsigma}]_k.
\end{align}
In order to apply the DA-MMSE receiver in \eqref{eq:da-mmse}, the BS should estimate the effective channel matrix ${\bf C}_{y\varsigma}$ and the received data signal correlation matrix ${\bf C}_{zz}+\sigma^2{\bf I}_M$. Since the signals  received at different antennas of the BS are independent for the channel model in \eqref{eq:gk}, the optimum effective channel estimation can be implemented element-wise. Using this, in this paper, we  present several schemes for effective channel estimation. However, the estimation of the received signal correlation matrix ${\bf C}_{zz}+\sigma^2{\bf I}_M$ involves the received pilot signals at all the BS antennas.   The received signal correlation matrix is conditioned on a channel realization and thus changes for each coherence block, hence it does not represent long-term statistics unlike the large-scale fading coefficients. This correlation matrix can be estimated using the data collected in each coherence block but the errors can be substantial due to the short data length especially when coherence length is small. As an alternative, we can simplify ${\bf C}_{\mu\mu}$ in \eqref{eq:da-mmse} as ${\bf C}_{\mu\mu}\odot{\bf I}_M$ and obtain the element-wise DA-MMSE (EW-DA-MMSE) receiver
\begin{align}\label{eq:ew-da-mmse}
{\bf v}_k^{\text{EW-DA-MMSE}}=\big({\bf C}_{\mu\mu}\odot{\bf I}_M+\sum_{i\neq k}[{\bf C}_{y\varsigma}]_i[{\bf C}_{y\varsigma}]_i^H\big)^{-1}[{\bf C}_{y\varsigma}]_k,
\end{align}
where the BS should estimate only the diagonal elements of ${\bf C}_{\mu\mu}$ using element-wise techniques which are more computationally efficient. Note that, the simplification ${\bf C}_{\mu\mu}\odot{\bf I}_M$ has been used in several papers for analytical tractability. In \cite{emil_hardware}, the effect of neglecting off-diagonal elements of distortion correlation matrix has been discussed. Here, we propose the receiver \eqref{eq:ew-da-mmse} for an efficient element-wise channel estimation from a different perspective, i.e., not to analyze its effect on SE, but to implement a practical receiver.  We will analyze the performance of this receiver compared to the DA-MMSE receiver that does not neglect off-diagonal elements in evaluating \eqref{eq:SINRk}. Distortion-aware maximum-ratio combining (DA-MRC) and regularized zero-forcing (DA-RZF) receivers can also be used without knowing ${\bf C}_{\mu\mu}$ but only the effective channels ${\bf C}_{y\varsigma}$:
\begin{align}
& {\bf v}_k^{\text{DA-MRC}}=[{\bf C}_{y\varsigma}]_k \label{eq:mrc}, \\
& {\bf v}_k^{\text{DA-RZF}}=[{\bf C}_{y\varsigma}\big({\bf C}_{y\varsigma}^H{\bf C}_{y\varsigma}+\sigma^2{\bf I}_K\big)^{-1}]_k \label{eq:rzf}. 
\end{align}

We will now look at the SE of the distortion-aware receivers using the 3GPP Urban Microcell model in \cite{3gpp} with a
2\,GHz carrier frequency and 20\,MHz bandwidth. The large-scale fading coefficients, shadowing parameters, probability of LOS, and the Rician factors are simulated based on \cite[Table B.1.2.1-1, B.1.2.1-2, B.1.2.2.1-4]{3gpp}. The BS antenna array and the UE heights from the ground are 10\,m and 1.5\,m, respectively. The noise variance is $\sigma^2=-96$\,dBm. The number of BS antennas is $M=100$, and $K=10$ users are uniformly distributed in a cell of 250\,m$\times$250\,m. The BS and UE hardware distortions are both modeled using a 3rd-order quasi-memoryless polynomial whose coefficients  are obtained by curve-fitting to the AM/AM and AM/PM distortions of a measured GaN amplifier operating at 2.1\,GHz; see \cite{ericsson}. The backoff parameters $b_{\text{off}}^{\text{BS}}$ and $b_{\text{off}}^{\text{UE}}$ are both 7\,dB.  Maximum transmission power for the UE antennas is 200\,mW and the heuristic uplink power control in \cite[Section 7.1.2]{emil_book} with $\Delta=20$\,dB is applied to determine $\{p_k\}$. 

\begin{figure}[t!]
	\centering
	\includegraphics[trim={3.8cm 0.35cm 6.7cm 3.4cm},clip,width=3.45in]{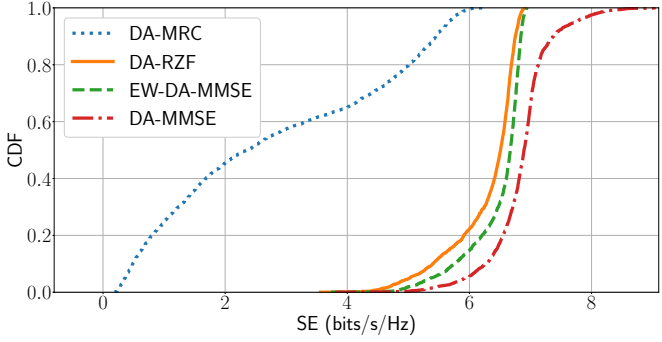}
	\caption{SE of DA-MRC, DA-RZF, EW-DA-MMSE, and DA-MMSE for $M=100$ and $K=10$.}
	\label{fig:se}
\end{figure}

Fig.~\ref{fig:se} shows the cumulative distribution function (CDF) of the SE of an arbitrary UE. The figure is generated from 250 different setups where the results of 100 channel realizations are averaged for each point. We observe that the SE of the EW-DA-MMSE receiver provides higher rate by exploiting the diagonal terms of the distortion correlation matrix compared to DA-MRC and DA-RZF which only use the effective channels. Although DA-MMSE results in higher SE, especially for high signal-to-noise ratio (SNR) UEs, EW-DA-MMSE allows using efficient  element-wise estimation techniques as we will see later. 

\section{Effective Channel Estimation with LMMSE \label{lmmse}}
In this section, we derive the LMMSE estimator of the effective channel. Let $\tau_{\text{p}}$ denote the uplink training duration in samples per coherence block. In the uplink training phase, all users simultaneously send pilot sequences to the BS. Let $\bm{\varphi}_k \in \mathbb{C}^{\tau_{\text{p}}}$ denote the pilot sequence of the $k^{\textrm{th}}$ UE where $||\bm{\varphi}_k||^2=\tau_{\text{p}}$, for $k=1,...,K$. Using the same hardware impairment model as in \eqref{eq:sk}, let $\tilde{\varphi}_{kn}$ denote the $n^\textrm{th}$ element of the UE hardware distorted pilot sequence, i.e.,  
\begin{align} \label{eq:pilot}
\tilde{\varphi}_{kn}=\tilde{b}_0\varphi_{kn}+\tilde{b}_1|\varphi_{kn}|^2\varphi_{kn}, \ \ \ n=1,\ldots,\tau_{\text{p}}.
\end{align}
The transmitted pilot sequence for the $k^\textrm{th}$ UE is given by $\{\sqrt{\tilde{\eta}_k}\tilde{\varphi}_{kn}\}$ where $\tilde{\eta}_k$ is selected as follows to make the average transmit power equal to $p_k$:
\begin{align} \label{eq:tilde-eta}
\tilde{\eta}_k=\frac{\tau_{\text{p}}p_k}{\sum_{n=1}^{\tau_{\text{p}}}|\tilde{\varphi}_{kn}|^2}.
\end{align} 
Let ${\bf z}^{\text{p}}_m \in \mathbb{C}^{\tau_{\text{p}}}$ denote the noise-free distorted signal at the $m^{\textrm{th}}$ antenna of the BS. The $n^{\textrm{th}}$ element of ${\bf z}^{\text{p}}_m $ is given by
\begin{align}\label{zpmn}
&z^{\text{p}}_{mn}=\sum_{l=0}^1\tilde{a}_{lm}\bigg(\sum_{k=1}^Kg_{km}\sqrt{\tilde{\eta}_k}\tilde{\varphi}_{kn}\bigg)\bigg|\sum_{k=1}^Kg_{km}\sqrt{\tilde{\eta}_k}\tilde{\varphi}_{kn}\bigg|^{2l}, \nonumber \\
&\hspace{0.8cm}\ \ n=1,...,\tau_{\text{p}}.
\end{align}
Then, the received baseband signal at the $m^{\textrm{th}}$ antenna of the BS in the uplink training phase is given by
\begin{align} \label{eq:ypm}
{\bf y}^{\text{p}}_{m}={\bf z}^{\text{p}}_m+{\bf n}^{\text{p}}_m,
\end{align}
where ${\bf n}^{\text{p}}_m\in \mathbb{C}^{\tau_{\text{p}}}$ is the uncorrelated thermal noise with ${\bf n}^{\text{p}}_m \sim \mathcal{N}_{\mathbb{C}}({\bf 0}_{\tau_{\text{p}}}, \sigma^2{\bf I}_{\tau_{\text{p}}})$. 

The ideal channel estimator is the MMSE estimator, which is normally used for distortion-free systems \cite{emil_book}. However, the MMSE channel estimator is hard to compute using the received signal in \eqref{eq:ypm} since it is not a linear Gaussian model unlike its distortion-free counterpart. Instead, we will restrict ourselves to the LMMSE estimator as a benchmark for the deep learning solution we will propose in Section \ref{deep}. Two different LMMSE channel estimation schemes can be designed under hardware impairments. The first one is a distortion-unaware LMMSE estimator which simply neglects the third order non-linear distortions and assumes an ideal linear model. In this case, the distortion-unaware LMMSE estimator estimates the physical channels $\{g_{km}\}$. The second option is to estimate the effective channel matrix presented in the previous section, ${\bf C}_{y\varsigma}$ from \eqref{eq:ypm}, and exploiting the expressions derived in previous sections. Now, we will discuss these approaches in detail.

\subsection{Distortion-Unaware LMMSE Estimator}
If we assume that the pilot vectors $\{\bm{\varphi}_k\}$ are mutually orthogonal, i.e. $\bm{\varphi}_k^H{\bm \varphi_{k^{\prime}}}=0$, $\forall k^{\prime}\neq k$, the distortion-unaware LMMSE estimate of the physical channel $g_{km}$,  which neglects the distortions ($\tilde{a}_{0m}=1$, $\tilde{a}_{1m}=0$, $\tilde{b}_0=1$, $\tilde{b}_{1}=0$, and $\tilde{\eta}_k=p_k$) is given by
\begin{align} \label{eq:dist-un}
&\hat{g}_{km}=\bar{ g}_{km}+\frac{\sqrt{p_k}\beta_k}{\tau_{\text{p}}p_k\beta_k+\sigma^2}(\bm{\varphi}_k^H{\bf y}_m^{\text{p}}-\sqrt{p_k}\tau_{\text{p}}\bar{g}_{km}).
\end{align}
This is the true LMMSE estimator in the absence of distortion, but can also be used by a BS unaware of its and the UEs' distortions.
\subsection{Distortion-Aware LMMSE Estimator}
The distortion-aware LMMSE estimator takes into account the first- and second-order statistics of the distorted signals and effective channel while effectively  treating the additive distortion term $\bm{\mu}$ in \eqref{eq:Bussgang} as a colored noise term which is independent of data signal vector $\bm{\varsigma}$, because the LMMSE estimator would coincide with the optimal MMSE estimator in that special case. However, $\bm{\mu}$ is clearly a function of $\bm{\varsigma}$ and more efficient methods can be developed to exploit the non-linear hardware characteristics, which is what we will do in Section \ref{deep}.

{\bf Remark:} The effective channels are functions of  not only the physical channels but also the statistics of the data signals, which makes the estimation of effective channels modulation-dependent unlike the physical channel estimation in ideal linear systems.

We consider LMMSE estimation of the effective channel matrix ${\bf C}_{y\varsigma}$ whose expression is given in \eqref{eq:eff-finite} for the data signals with the $90^{\circ}$ circular shift property. The LMMSE estimate of the $(m,k)$th element of  ${\bf C}_{y\varsigma}$ given ${\bf y}^{\text{p}}_m$ is given by
\begin{align} \label{eq:dist-aw}
&[\hat{\bf C}_{y\varsigma}]_{mk}=[\bar{\bf C}_{y\varsigma}]_{mk}+{\bf C}_{[C_{y\varsigma}]_{mk}y^{\text{p}}_m}{\bf C}_{y^{\text{p}}_my^{\text{p}}_m}^{-1}({\bf y}^{\text{p}}_m-\bar{\bf y}^{\text{p}}_m),\nonumber \\
&k=1,....,K, \ \ \ m=1,...,M,
\end{align}
where 
\begin{align} 
&\bar{\bf y}^{\text{p}}_m=\mathbb{E}\{{\bf y}^{\text{p}}_m\}\in \mathbb{C}^{\tau_{\text{p}}} \label{eq:y-mean} ,\\
&\bar{\bf C}_{y\varsigma}=\mathbb{E}\{{\bf C}_{y\varsigma}\}\in \mathbb{C}^{M\times K} \label{eq:eff-mean}, \\
&{\bf C}_{[C_{y\varsigma}]_{mk}y^{\text{p}}_m}=\mathbb{E}\bigg\{\big([{\bf C}_{y\varsigma}]_{mk}-[\bar{\bf C}_{y\varsigma}]_{mk}\big)\big({\bf y}^{\text{p}}_m-\bar{\bf y}^{\text{p}}_m\big)^H\bigg\}\nonumber\\
&\hspace{2cm}\in \mathbb{C}^{1\times \tau_{\text{p}}} \label{eq:y-eff-cov}, \\
&{\bf C}_{y^{\text{p}}_my^{\text{p}}_m}=\mathbb{E}\bigg\{\big({\bf y}^{\text{p}}_m-\bar{\bf y}^{\text{p}}_m\big)\big({\bf y}^{\text{p}}_m-\bar{\bf y}^{\text{p}}_m\big)^H\bigg\}\in\mathbb{C}^{\tau_{\text{p}} \times \tau_{\text{p}}} \label{eq:y-cov}.
\end{align}
We will now compute the expectations in \eqref{eq:y-mean}-\eqref{eq:y-cov}. Let us first define the following vectors:
\begin{align}
\widetilde{\bm{\varphi}}_n=&[ \ \sqrt{\beta_1\tilde{\eta}_1}\tilde{\varphi}_{1n}^{*} \ \ldots  \ \sqrt{\beta_K\tilde{\eta}_K}\tilde{\varphi}_{Kn}^{*} \ ]^T\in \mathbb{C}^{K}, \nonumber\\
&n=1,\ldots,\tau_{\text{p}} \label{eq:varphi-tilde}, \\
\bar{\bf h}_m=&[ \ \bar{g}_{1m}/\sqrt{\beta}_1 \ \ldots \ \bar{g}_{Km}/\sqrt{\beta_K} \ ]^T\in \mathbb{C}^{K}, \nonumber\\
&m=1,\ldots,M, \label{eq:h-bar} \\
\widetilde{\bf h}_m=&[ \ (g_{1m}-\bar{g}_{1m})/\sqrt{\beta}_1 \ \ldots \ (g_{Km}-\bar{g}_{Km})/\sqrt{\beta_K} \ ]^T\nonumber\\
&\in \mathbb{C}^{K}, \ \ \ m=1,\ldots,M. \label{eq:h-tilde} 
\end{align}
Note that the elements of $\widetilde{\bf h}_m$ are i.i.d. $\mathcal{N}_{\mathbb{C}}(0,1)$, hence we can use the results of Lemma 2 when computing the expectations. Using \eqref{eq:varphi-tilde}-\eqref{eq:h-tilde}, we can express 
\begin{align}
y^{\text{p}}_{mn}=&\tilde{a}_{0m}\widetilde{\bm{\varphi}}_n^H(\bar{\bf h}_m+\widetilde{\bf h}_m)+\tilde{a}_{1m}\widetilde{\bm{\varphi}}_n^H(\bar{\bf h}_m+\widetilde{\bf h}_m)\times \nonumber\\&(\bar{\bf h}_m+\widetilde{\bf h}_m)^H\widetilde{\bm{\varphi}}_n
\widetilde{\bm{\varphi}}_n^H(\bar{\bf h}_m+\widetilde{\bf h}_m)+n^{\text{p}}_{mn} \label{eq:ypmn}, \\
[{\bf C}_{y\varsigma}]_{mk}=&\tilde{c}_{0m}{\bf e}_k^H(\bar{\bf h}_m+\widetilde{\bf h}_m)+\tilde{c}_{1m}{\bf e}_k^H(\bar{\bf h}_m+\widetilde{\bf h}_m)\times\nonumber\\
&(\bar{\bf h}_m+\widetilde{\bf h}_m)^H{\bf e}_k{\bf e}_k^H(\bar{\bf h}_m+\widetilde{\bf h}_m) \nonumber \\
& +\tilde{c}_{2m}\sum_{l\neq k}^K{\bf e}_l^H(\bar{\bf h}_m+\widetilde{\bf h}_m)(\bar{\bf h}_m+\widetilde{\bf h}_m)^H{\bf e}_l\times\nonumber\\
&\hspace{1.7cm}{\bf e}_k^H(\bar{\bf h}_m+\widetilde{\bf h}_m) \label{eq:eff-new},
\end{align}
where ${\bf e}_k\in \mathbb{C}^{K}$ is the vector whose only nonzero element is $\sqrt{\beta_k\eta_k}$ at the index $k$ and the following parameters are defined for ease of notation:
\begin{align}
\tilde{c}_{0m}=&\tilde{a}_{0m}(\tilde{b}_0+\zeta_4\tilde{b}_1), \label{eq:tilde-c0}\\
\tilde{c}_{1m}=&\tilde{a}_{1m}\big(\zeta_{10}B_{1,1,1}+2\zeta_8B_{1,1,0}+\zeta_8B_{1,0,1}+2\zeta_6B_{0,0,1}\nonumber\\
&+\zeta_6B_{0,1,0}+\zeta_4B_{0,0,0}\big), \label{eq:tilde-c1}\\
\tilde{c}_{2m}=&2\tilde{a}_{1m}(\tilde{b}_{0}+\zeta_4\tilde{b}_{1})(\zeta_6B_{1,1}+\zeta_4B_{1,0}+\zeta_4B_{0,1}+B_{0,0}). \label{eq:tilde-c2}
\end{align}
{\color{blue} Let us define the following functions for ease of notation:
	\begin{align}
&\mathcal{E}_{m1}({\bf a}_1,{\bf b}_1)\triangleq \mathbb{E}\big\{{\bf a}_1^H(\bar{\bf h}_m+\widetilde{\bf h}_m)(\bar{\bf h}_m+\widetilde{\bf h}_m)^H{\bf b}_1\big\}, \label{eq:exp_func1} \\
&\mathcal{E}_{m2}({\bf a}_1,{\bf b}_1,{\bf a}_2,{\bf b}_2)\triangleq \mathbb{E}\big\{{\bf a}_1^H(\bar{\bf h}_m+\widetilde{\bf h}_m)\times\nonumber\\
&\hspace{1cm}(\bar{\bf h}_m+\widetilde{\bf h}_m)^H{\bf b}_1{\bf a}_2^H(\bar{\bf h}_m+\widetilde{\bf h}_m)(\bar{\bf h}_m+\widetilde{\bf h}_m)^H{\bf b}_2\big\}, \label{eq:exp_func2} \\
&\mathcal{E}_{m3}({\bf a}_1,{\bf b}_1,{\bf a}_2,{\bf b}_2,{\bf a}_3,{\bf b}_3)\triangleq \mathbb{E}\big\{{\bf a}_1^H(\bar{\bf h}_m+\widetilde{\bf h}_m)\times\nonumber\\
&\hspace{1cm}(\bar{\bf h}_m+\widetilde{\bf h}_m)^H{\bf b}_1{\bf a}_2^H(\bar{\bf h}_m+\widetilde{\bf h}_m)(\bar{\bf h}_m+\widetilde{\bf h}_m)^H{\bf b}_2\times\nonumber\\
&\hspace{1cm}{\bf a}_3^H(\bar{\bf h}_m+\widetilde{\bf h}_m)(\bar{\bf h}_m+\widetilde{\bf h}_m)^H{\bf b}_3\big\} \label{eq:exp_func3}.
	\end{align}
}Now, using the definitions in \eqref{eq:varphi-tilde}-{\color{blue}\eqref{eq:exp_func3}}, the elements of \eqref{eq:y-mean}-\eqref{eq:y-cov} are given in (\ref{eq:y-bar-n})-(\ref{eq:y-cov-new}) at the top of the following page.
\begin{figure*}
\begin{align}
&\bar{y}^{\text{p}}_{mn}=\mathbb{E}\{y^{\text{p}}_{mn}\}=\tilde{a}_{0m}\widetilde{\bm{\varphi}}_n^H\bar{\bf h}_m+\tilde{a}_{1m}\widetilde{\bm{\varphi}}_n^H\bar{\bf h}_m\bar{\bf h}_m^H\widetilde{\bm{\varphi}}_n\widetilde{\bm{\varphi}}_n^H\bar{\bf h}_m+2\tilde{a}_{1m}\widetilde{\bm{\varphi}}_n^H\bar{\bf h}_m\widetilde{\bm{\varphi}}_n^H\widetilde{\bm{\varphi}}_n \label{eq:y-bar-n}, \\
& [\bar{\bf C}_{y\varsigma}]_{mk}=\mathbb{E}\{[{\bf C}_{y\varsigma}]_{mk}\}=\tilde{c}_{0m}{\bf e}_k^H\bar{\bf h}_m+\tilde{c}_{1m}{\bf e}_k^H\bar{\bf h}_m\bar{\bf h}_m^H{\bf e}_k{\bf e}_k^H\bar{\bf h}_m+2\tilde{c}_{1m}{\bf e}_k^H\bar{\bf h}_m{\bf e}_k^H{\bf e}_k+\tilde{c}_{2m}\sum_{l\neq k}^K{\bf e}_l^H\bar{\bf h}_m\bar{\bf h}_m^H{\bf e}_l{\bf e}_k^H\bar{\bf h}_m\nonumber\\
&\hspace{3.6cm}+\tilde{c}_{2m}\sum_{l\neq k}^K{\bf e}_l^H\bar{\bf h}_m{\bf e}_k^H{\bf e}_l+\tilde{c}_{2m}\sum_{l\neq k}^K{\bf e}_l^H{\bf e}_l{\bf e}_k^H\bar{\bf h}_m \nonumber \\
& \hspace{3.6cm} =\tilde{c}_{0m}\sqrt{\eta_k}\bar{g}_{km}+\tilde{c}_{1m}\sqrt{\eta_k}\bar{g}_{km}\big(\eta_k|\bar{g}_{km}|^2+2\eta_k\beta_k\big)+\tilde{c}_{2m}\sqrt{\eta_k}\bar{g}_{km}\sum_{l\neq k}^K\big(\eta_l|\bar{g}_{lm}|^2+\eta_l\beta_l\big) \label{eq:eff-new-bar},\\
&\big[{\bf C}_{[C_{y\varsigma}]_{mk}y^{\text{p}}_m}\big]_n=\mathbb{E}\big\{\big([{\bf C}_{y\varsigma}]_{mk}-[\bar{\bf C}_{y\varsigma}]_{mk}\big)\big({y}^{\text{p}}_{mn}-\bar{y}^{\text{p}}_{mn}\big)^{*}\big\}=\tilde{c}_{0m}\tilde{a}_{0m}^{*}\mathcal{E}_{m1}({\bf e}_k,\widetilde{\bm{\varphi}}_n)+\tilde{c}_{0m}\tilde{a}_{1m}^{*}\mathcal{E}_{m2}({\bf e}_k,\widetilde{\bm{\varphi}}_n,\widetilde{\bm{\varphi}}_n,\widetilde{\bm{\varphi}}_n)\nonumber\\
&\hspace{3.6cm}+\tilde{c}_{1m}\tilde{a}_{0m}^{*}\mathcal{E}_{m2}({\bf e}_k,{\bf e}_k,{\bf e}_k,\widetilde{\bm{\varphi}}_n)+\tilde{c}_{1m}\tilde{a}_{1m}^{*}\mathcal{E}_{m3}({\bf e}_k,{\bf e}_k,{\bf e}_k,\widetilde{\bm{\varphi}}_n,\widetilde{\bm{\varphi}}_n,\widetilde{\bm{\varphi}}_n)\nonumber\\
&\hspace{3.6cm}+\tilde{c}_{2m}\tilde{a}_{0m}^{*}\sum_{l\neq k}^K\mathcal{E}_{m2}({\bf e}_l,{\bf e}_l,{\bf e}_k,\widetilde{\bm{\varphi}}_n)+\tilde{c}_{2m}\tilde{a}_{1m}^{*}\sum_{l\neq k}^K\mathcal{E}_{m3}({\bf e}_l,{\bf e}_l,{\bf e}_k,\widetilde{\bm{\varphi}}_n,\widetilde{\bm{\varphi}}_n,\widetilde{\bm{\varphi}}_n)-[\bar{\bf C}_{y\varsigma}]_{mk}\big(\bar{y}^{\text{p}}_{mn}\big)^{*}  \label{eq:y-eff-cov-new}, \\
&\big[{\bf C}_{y^{\text{p}}_my^{\text{p}}_m}\big]_{nj}=\mathbb{E}\big\{\big({ y}^{\text{p}}_{mn}-\bar{y}^{\text{p}}_{mn}\big)\big({y}^{\text{p}}_{mj}-\bar{y}^{\text{p}}_{mj}\big)^{*}\big\}=|\tilde{a}_{0m}|^2\mathcal{E}_{m1}(\widetilde{\bm{\varphi}}_n,\widetilde{\bm{\varphi}}_j)+\tilde{a}_{0m}\tilde{a}_{1m}^{*}\mathcal{E}_{m2}(\widetilde{\bm{\varphi}}_n,\widetilde{\bm{\varphi}}_j,\widetilde{\bm{\varphi}}_j,\widetilde{\bm{\varphi}}_j)\nonumber\\
&\hspace{3.6cm}+\tilde{a}_{1m}\tilde{a}_{0m}^{*}\mathcal{E}_{m2}(\widetilde{\bm{\varphi}}_n,\widetilde{\bm{\varphi}}_n,\widetilde{\bm{\varphi}}_n,\widetilde{\bm{\varphi}}_j)+|\tilde{a}_{1m}|^2\mathcal{E}_{m3}(\widetilde{\bm{\varphi}}_n,\widetilde{\bm{\varphi}}_n,\widetilde{\bm{\varphi}}_n,\widetilde{\bm{\varphi}}_j,\widetilde{\bm{\varphi}}_j,\widetilde{\bm{\varphi}}_j)-\bar{y}^{\text{p}}_{mn}\big(\bar{y}^{\text{p}}_{mj}\big)^{*}+\sigma^2\delta_{nj} . \label{eq:y-cov-new} 
\end{align}
\hrulefill
\end{figure*}
Note that most of the terms in \eqref{eq:lemma2-2}-\eqref{eq:lemma2-4} become zero since the vector $\widetilde{\bf h}_m$ is Gaussian distributed. In addition, the circular symmetric property of $\widetilde{\bf h}_m$ results in zero expectations for some terms in \eqref{eq:y-eff-cov-new} and \eqref{eq:y-cov-new}. For ease of notation, let us define the following set of functions:
\begin{align}\label{eq:Hm}
H_m({\bf x},{\bf y})\triangleq{\bf x}^H\bar{\bf h}_m\bar{\bf h}_m^H{\bf y}, \ \ m=1,\ldots,M. 
\end{align}
Using Lemma 2 and the functions in \eqref{eq:Hm}, we can obtain the following lemma for the calculation of the expectations in \eqref{eq:y-eff-cov-new} and \eqref{eq:y-cov-new}.

\emph{Lemma 3:} Let ${\bf a}_1\in \mathbb{C}^{K}$, ${\bf a}_2 \in \mathbb{C}^{K}$, ${\bf a}_3 \in \mathbb{C}^{K}$, ${\bf b}_1 \in \mathbb{C}^{K}$, ${\bf b}_2 \in \mathbb{C}^{K}$, and ${\bf b}_3 \in \mathbb{C}^{K}$ denote arbitrary deterministic vectors. For the vectors defined in \eqref{eq:h-bar}-\eqref{eq:h-tilde}, the following hold: 
{\color{blue}\begin{align}
&\text{1) } \mathcal{E}_{m1}({\bf a}_1,{\bf b}_1)=H_m({\bf a}_1,{\bf b}_1)+{\bf a}_1^H{\bf b}_1 \label{lemma3-1}. \\
&\text{2) } \mathcal{E}_{m2}({\bf a}_1,{\bf b}_1,{\bf a}_2,{\bf b}_2)=H_m({\bf a}_1,{\bf b}_1)H_m({\bf a}_2,{\bf b}_2)\nonumber \\
&\hspace{0.6cm} +\sum_{\substack{i_1, i_2, j_1, j_2 \in \{1,2\}\\
i_1\neq j_1, \ i_2 \neq j_2 \ }}\Bigg(H_m({\bf a}_{i_1},{\bf b}_{i_2})+\frac{{\bf a}_{i_1}^H{\bf b}_{i_2}}{2}\Bigg){\bf a}_{j_1}^H{\bf b}_{j_2} \label{lemma3-2}. \\
&\text{3) } \mathcal{E}_{m3}({\bf a}_1,{\bf b}_1,{\bf a}_2,{\bf b}_2,{\bf a}_3,{\bf b}_3)\nonumber \\
&\hspace{0.0cm}=H_m({\bf a}_1,{\bf b}_1)H_m({\bf a}_2,{\bf b}_2)H_m({\bf a}_3,{\bf b}_3)\nonumber \\
&\hspace{0.0cm}+\sum_{\substack{i_1, i_2, j_1, j_2, k_1, k_2 \in \{1,2\}\\
		i_1\neq j_1 \neq k_1, \ i_2 \neq j_2 \neq k_2, \\ i_1 \neq k_1, \ i_2 \neq k_2 }}
 \Bigg(\frac{H_m({\bf a}_{i_1},{\bf b}_{i_2})H_m({\bf a}_{j_1},{\bf b}_{j_2}){\bf a}_{k_1}^H{\bf b}_{k_2}}{4}\nonumber\\
&+\frac{H_m({\bf a}_{i_1},{\bf b}_{i_2}){\bf a}_{j_1}^H{\bf b}_{j_2}{\bf a}_{k_1}^H{\bf b}_{k_2}}{2}+\frac{{\bf a}_{i_1}^H{\bf b}_{i_2}{\bf a}_{j_1}^H{\bf b}_{j_2}{\bf a}_{k_1}^H{\bf b}_{k_2}}{6}\Bigg) \label{lemma3-3}.
\end{align}}
\begin{IEEEproof} This can easily be proved by expanding the products in the expectations and eliminating the terms with zero-mean by utilizing the circular symmetric property of $\widetilde{\bf h}_m$. For high-order moments, Lemma 2 is applied for the vectors $\widetilde{\bf h}_m$  whose elements are zero-mean unit-variance i.i.d. circularly symmetric Gaussian random variables. Note that the results of Lemma 3 follow considering all the combinations for the nonzero mean terms.
	\end{IEEEproof}

We can use these closed-form expressions to compute the distortion-aware LMMSE estimator in \eqref{eq:dist-aw}. However, finding the LMMSE estimator for the diagonal elements of the distortion correlation matrix ${\bf C}_{\mu\mu}$ in \eqref{eq:dist-corr} is very complicated. In the numerical results, we will use Monte Carlo estimation for these correlation elements and compare the performance of it with the proposed deep learning estimator.

In the next part, we will propose a deep learning based architecture for efficient estimation of the effective channels in \eqref{eq:eff-finite} and diagonal elements of distortion correlation matrix given in \eqref{eq:dist-corr}. It can both reduce complexity and improve estimation performance.
\section{Effective Channel and Element-Wise Distortion Correlation Estimation with Deep Learning \label{deep}}

In this section, we propose two deep feedforward neural networks with fully-connected layers in order to realize estimation of the effective channel and distortion correlation whose analytical expressions given in \eqref{eq:eff-finite}, \eqref{eq:dist-corr}, and \eqref{eq:Czz} are used to train the model-driven networks. 

A feedforward neural network with $P$ fully-connected layers presents a non-linear mapping from an input vector ${\bf r}_0\in\mathbb{R}^{N_0}$ to an output vector ${\bf r}_P\in \mathbb{R}^{N_P}$ through $P$ iterative functions:
\begin{align}\label{eq:deep-r}
{\bf r}_p=\sigma_p({\bf W}_p{\bf r}_{p-1}+{\bf b}_p), \ \ \ p=1,...,P,
\end{align}
where ${\bf W}_p\in \mathbb{R}^{N_p\times N_{p-1}}$ is the weighting matrix at the $p^{th}$ layer and ${\bf b}_p \in \mathbb{R}^{N_p}$ is the corresponding bias vector. $\sigma_p(.)$ is the activation function for the $p^{th}$ layer and it is used to introduce non-linearity to the considered mapping. Without the non-linearity, the overall mapping from the input vector to the output vector is simply an affine function. The power of deep learning lies in the use of effective non-linear activation functions in multiple successive layers. In this way, a properly designed deep learning network can learn how the hardware has impaired the desired signal during uplink training and data transmission. Furthermore, it can exploit this information to learn a more effective channel and distortion correlation estimation approach compared to the LMMSE-based methods derived in Section \ref{lmmse}.   In supervised learning, deep neural networks are trained using training data that is given by a set of input-output vector pairs, i.e., $\{{\bf r}^t_0,\tilde{\bf r}^t_P\}_{t=1}^T$ where $T$ is the training size. Here, $\tilde{\bf r}^t_P$ is the desired output for the given input vector ${\bf r}^t_0$. A loss function is used  for the optimization of the parameters $\{{\bf W}_p,{\bf b}_p\}_{p=1}^P$ as follows:
\begin{align}\label{eq:deep-loss}
L\big(\{{\bf W}_p,{\bf b}_p\}_{p=1}^P\big)=\frac{1}{T}\sum_{t=1}^Tl(\tilde{\bf r}^t_P,{\bf r}^t_P),
\end{align} 
where $l(.,.):\mathbb{R}^{N_P} \times \mathbb{R}^{N_P} \rightarrow \mathbb{R}$ is the loss function of the desired output and the actual output when ${\bf r}^t_0$ is the input. The deep learning optimization algorithms aim at minimizing the loss in \eqref{eq:deep-loss}. For further details on deep learning, please refer to the references \cite{deep_book},\cite{deep_physical}.

\begin{figure*}[t!]
	\begin{center}
		\begin{overpic}[width=12cm,tics=10]{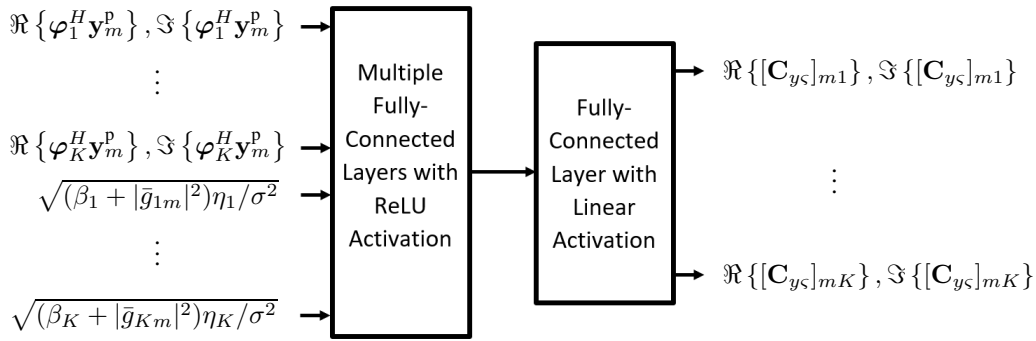}
			\put(-4,37.8){$\Re\left\{\bm{\varphi}_1^H{\bf y}_m^{\text{p}}\right\}, \Im\left\{\bm{\varphi}_1^H{\bf y}_m^{\text{p}}\right\}$}
				\put(12,31){$\vdots$}
		\put(-4,24.4){$\Re\left\{\bm{\varphi}_K^H{\bf y}_m^{\text{p}}\right\}, \Im\left\{\bm{\varphi}_K^H{\bf y}_m^{\text{p}}\right\}$}
	\put(-1,18.8){$\sqrt{(\beta_1+|\bar{g}_{1m}|^2)\eta_1/\sigma^2}$}
	\put(12,12){$\vdots$}
		\put(-4,5.6){$\sqrt{(\beta_K+|\bar{g}_{Km}|^2)\eta_K/\sigma^2}$}
		\put(75,32.6){$\Re\left\{[{\bf C}_{y\varsigma}]_{m1}\right\}, \Im\left\{[{\bf C}_{y\varsigma}]_{m1}\right\}$}
		\put(87,20){$\vdots$}
		\put(75,10){$\Re\left\{[{\bf C}_{y\varsigma}]_{mK}\right\}, \Im\left\{[{\bf C}_{y\varsigma}]_{mK}\right\}$}
			\end{overpic}
	\end{center}
	\caption{Deep feedforward neural network for effective channel estimation.} 
	\label{fig:deep1} 
\end{figure*}

\begin{figure*}[t!]
	\begin{center}
		\begin{overpic}[width=9cm,tics=10]{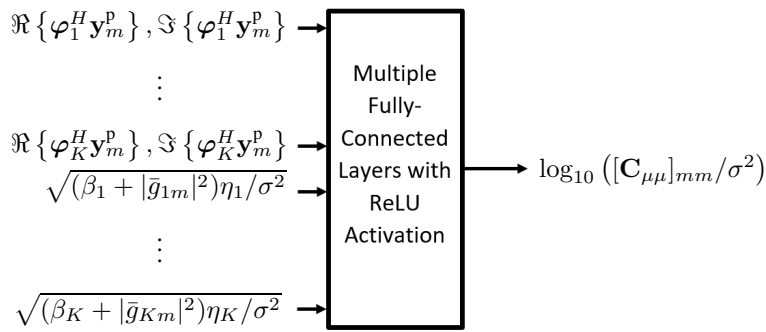}
	\put(-6,48.8){$\Re\left\{\bm{\varphi}_1^H{\bf y}_m^{\text{p}}\right\}, \Im\left\{\bm{\varphi}_1^H{\bf y}_m^{\text{p}}\right\}$}
\put(15.5,39){$\vdots$}
\put(-6,31.2){$\Re\left\{\bm{\varphi}_K^H{\bf y}_m^{\text{p}}\right\}, \Im\left\{\bm{\varphi}_K^H{\bf y}_m^{\text{p}}\right\}$}
\put(-1.1,25){$\sqrt{(\beta_1+|\bar{g}_{1m}|^2)\eta_1/\sigma^2}$}
\put(15.5,15){$\vdots$}
\put(-5.6,6.8){$\sqrt{(\beta_K+|\bar{g}_{Km}|^2)\eta_K/\sigma^2}$}
\put(72,27.6){$\log_{10}\left([{\bf C}_{\mu\mu}]_{mm}/\sigma^2\right)$}
\end{overpic}
\end{center}
	\caption{Deep feedforward neural network for diagonal elements of distortion correlation matrix.}
	\label{fig:deep2}
\end{figure*}

We propose the feedforward neural network structures in Fig.~\ref{fig:deep1} and Fig.~\ref{fig:deep2} for the estimation of effective channel and diagonal elements of the distortion correlation matrix. Since the small-scale fading coefficients are independent for each antenna of the BS, we will train the neural networks in Fig.~\ref{fig:deep1} and Fig.~\ref{fig:deep2} for a single antenna element and use it for each antenna for the estimation of effective channels and element-wise distortion correlation. Even if the small-scale fading coefficients are correlated, we can use these structures for a simple and computationally efficient approach since it is not mandatory to utilize the correlation. The elements of the effective channel and distortion correlation matrix are given in \eqref{eq:eff-finite} and \eqref{eq:dist-corr}, respectively. If we focus on the $m^\textrm{th}$ antenna's channels, the first $2K$ inputs of the proposed networks are the real and imaginary parts of the processed  received signals in uplink training phase by correlating them with pilot sequences as
\begin{align} \label{eq:first-inputs}
\bm{\varphi}_k^H{\bf y}_m^{\text{p}}, \ \ \ k=1,...,K, 
\end{align} 
which represents  a naive estimate  of $g_{km}$ without taking into account the additional distortion terms. 

{\bf Remark:} Note that we assumed orthogonal pilot sequences when deriving the distortion-unaware LMMSE estimator in \eqref{eq:dist-un} whereas we did not specify any structure for the pilot sequences in the distortion-aware LMMSE estimator in \eqref{eq:dist-aw}. Even if we use orthogonal pilot sequences, perfect despreading of the received signals by correlation in \eqref{eq:first-inputs}  is not possible unlike the distortion-free scenario. This means that the processed signals in \eqref{eq:first-inputs} are not independent for different users.

The other inputs of the neural networks are the square roots of the scaled channel gain over noise, i.e., $\sqrt{(\beta_k+|\bar{g}_{km}|^2)\eta_k/\sigma^2}$ for $k=1,\ldots,K$, which depend on the long-term channel parameters and known at the BS.

Note that the ReLU activation function \cite{deep_book,deep_physical} is used in the hidden layers of the deep neural networks presented in Fig.~\ref{fig:deep1} and Fig.~\ref{fig:deep2}.

The outputs of the channel estimator in Fig.~\ref{fig:deep1} are the real and imaginary parts of the effective channel elements. In the element-wise distortion correlation estimator in Fig.~\ref{fig:deep2}, we take the logarithm of the diagonal elements of the distortion correlation matrix after normalizing it with the noise variance. Note that  $[{\bf C}_{\mu\mu}]_{mm}/\sigma^2$ is always greater than or equal to $1$, hence the logarithm always results in a non-negative number. The reason for taking the logarithm is to make the distribution of the output more uniform, which improves the learning.

At the output layer of the deep neural network in Fig.~\ref{fig:deep1}, linear activation is used since the outputs can take both positive and negative values whereas the ReLU activation is used at the output layer in Fig.~\ref{fig:deep2} where we exploit the knowledge that the logarithm of the normalized diagonal elements of the distortion correlation matrix is always nonnegative.

When training the neural networks in Fig.~\ref{fig:deep1} and Fig.~\ref{fig:deep2}, one of the main difficulties is the fluctuant SNR values. In order to simplify the learning, we can arrange the order of  inputs and outputs such that their indices are according to descending or ascending channel gains which are the last $K$ inputs of the networks. It is observed empirically that this method improves the learning.

\section{Effective Channels for General Quasi-Memoryless Distortion and Deep Learning-Based Estimation \label{general}}

We will now derive the effective channel during data transmission for general quasi-memoryless distortion of any order at the BS and UEs.

If we assume $(2R+1)^{\textrm{th}}$ order quasi-memoryless distortion at the UEs, the transmitted distorted signal from the $k^{\textrm{th}}$ UE is $s_k=\sqrt{\eta_k}\upsilon_k$ where
\begin{align} \label{eq:upsilon}
&\upsilon_k=\sum_{r=0}^R\tilde{b}_r|\varsigma_k|^{2r}\varsigma_k,
\end{align}
where $\tilde{b}_r$ is given by
\begin{align}\label{eq:bk2}
&\tilde{b}_r=\frac{b_r}{(b_{\text{off}}^{\text{UE}})^r}, \ \ r=0,1,\ldots,R,
\end{align}
and $\{b_r\}$ are the reference polynomial coefficients consistent with \eqref{eq:bk}. The following lemma proves an important result that we will use later on.

\emph{Lemma 4:} For zero-mean data symbols $\varsigma_k$ satisfying  
\begin{align}\label{eq:lemma4-1}
\mathbb{E}\{\varsigma_k^{l_1}(\varsigma_k^{*})^{l_2}\}=0, \ \ l_1-l_2\neq 4i \ \ \text{for any } i\in \mathbb{Z},
\end{align}
for any $l_1,l_2\in \mathbb{Z}^{+}$, it is true for  the distorted data symbol $\upsilon_k$ defined in \eqref{eq:upsilon} that
\begin{align}
&\mathbb{E}\{\upsilon_k^{l_1}(\upsilon_k^{*})^{l_2-1}\varsigma_k^{*}\}=0, \ \ l_1-l_2\neq 4i \ \ \text{for any } i\in \mathbb{Z}, \label{eq:lemma4-2a} \\
&\mathbb{E}\{\upsilon_k^{l_1}(\upsilon_k^{*})^{l_2}\}=0, \ \ l_1-l_2\neq 4i \ \ \text{for any } i\in \mathbb{Z}, \label{eq:lemma4-2b}
\end{align}
for any $l_1,l_2\in \mathbb{Z}^{+}$.
\begin{IEEEproof} The proof easily follows from the definition of $\upsilon_k$ in \eqref{eq:upsilon}.
\end{IEEEproof}

Generalizing the notation and analysis from Section \ref{system} to $(2T+1)^{\textrm{th}}$ order quasi-memoryless distortion at the BS, the noise-free distorted digital baseband signal at BS antenna $m$ during uplink data transmission phase is given by
\begin{align} \label{eq:zm}
z_m=\sum_{t=0}^T\tilde{a}_{tm}|u_m|^{2t}u_m, \ \ m=1,\ldots,M,
\end{align}
where $\{\tilde{a}_{tm}\}$ are the distortion polynomial coefficients as defined in \eqref{eq:tilde-alm}. Then, the $(m,k)$th element of the effective channel ${\bf C}_{y\varsigma}$, i.e., $[{\bf C}_{y\varsigma}]_{mk}$ is given by
\begin{align} \label{eq:C_ysigma}
[{\bf C}_{y\varsigma}]_{mk}&=\mathbb{E}_{|{\bf G}}\{y_m\varsigma_k^{*}\}=\mathbb{E}_{|{\bf G}}\{z_m\varsigma_k^{*}\}\nonumber\\
&=\sum_{t=0}^T\tilde{a}_{tm}\mathbb{E}_{|{\bf G}}\{|u_m|^{2t}u_m\varsigma_k^{*}\}.
\end{align}
For data signals satisfying the $90^{\circ}$ circular shift symmetry, if we define $S_t=\min(t+1,K)$,  $\mathbb{E}_{|{\bf G}}\{|u_m|^{2t}u_m\varsigma_k^{*}\}$ in \eqref{eq:C_ysigma} can be expressed as in \eqref{eq:moments} at the top of the following page.
\begin{figure*}
\begin{align} \label{eq:moments}
&\mathbb{E}_{|{\bf G}}\{|u_m|^{2t}u_m\varsigma_k^{*}\}=\mathbb{E}_{|{\bf G}}\bigg\{\bigg(  \sum_{l=1}^K\tilde{g}_{lm}\upsilon_l\bigg)^{t+1}\bigg(  \sum_{l=1}^K\tilde{g}_{lm}^{*}\upsilon_l^{*}\bigg)^t\varsigma_k^{*}\bigg\} \nonumber \\
&=\sum_{\substack{k_1,\ldots,k_{S_t}, l_1,\ldots,l_{S_t}\\k_1+k_2+\ldots+k_{S_t}=t+1,\\l_1+l_2+\ldots+l_{S_t}=t+1, \\ k_s-l_s=4i \text{ for some integer} \ i \text{ for } s=1,\ldots,S_t, \\ l_1\geq 1, \\ k_{S_t}\geq k_{S_t-1} \geq \ldots \geq k_2, \\ l_s\geq l_{s-1} \text{ if } k_s=k_{s-1} \text{ for } s=3,\ldots,S_t}}  \binom{t+1}{k_1,k_2,\ldots,k_{S_t}} \binom{t}{l_1-1,l_2,\ldots,l_{S_t}}  \mathbb{E}\{\upsilon^{k_1}(\upsilon^{*})^{l_1-1}\varsigma^{*}\}\times \nonumber \\&\Bigg(\prod_{s=2}^{S_t}\mathbb{E}\{\upsilon^{k_s}(\upsilon^{*})^{l_s}\}\Bigg) {\tilde{g}_{km}}^{k_1}({\tilde{g}_{km}}^{*})^{l_1-1} \sum_{\substack{f_2,\ldots,f_{S_t}\\ f_i\neq f_j \text{ for } i\neq j\\f_i\neq k \text{ for } i=2,\ldots,S_t\\ f_i>f_j \text{ if } k_i=k_j \text{ and } l_i=l_j \text{ for } i> j} }\prod_{i=2}^{S_t}\tilde{g}_{f_im}^{k_i}({\tilde{g}_{f_im}}^{*})^{l_i}, \nonumber \\
&t=0,\ldots,T, \ \ \ m=1,\ldots,M, \ \ \ k=1,\ldots,K.
\end{align}
\hrulefill
\end{figure*}
Note that $\eqref{eq:moments}$ is derived using some combinatorial manipulations. The conditions under the  summation symbols ensure that all the terms in \eqref{eq:moments} are distinct. Furthermore, most of the terms become zero due to the conditions by Lemma 4.  Even though \eqref{eq:moments} may seem complex, $\mathbb{E}_{|{\bf G}}\{|u_m|^{2t}u_m\varsigma_k^{*}\}$ can be calculated easily for small $t$ values. Note that $t$ is at most $T$, which is typically $1,2,3$, or $4$ when dealing with non-linear hardware \cite{book_rf}, \cite{ericsson}. 

Note that the above analytical results can be efficiently used to generate large number of training samples for the deep learning network in Fig.~\ref{fig:deep1} for the effective channel estimation. Since it is hard to derive the elements of distortion correlation matrix ${\bf C}_{\mu\mu}$ for general-order non-linear model, we restrict ourselves to DA-MRC and DA-RZF receivers in \eqref{eq:mrc}, \eqref{eq:rzf} which use only the effective channel estimates for signal detection under general hardware distortion.

\section{Numerical Results \label{numerical}}

In this section, we compare the estimation performance of the proposed deep-learning-based estimators with several benchmarks. 
The polynomial coefficients of the distortion model in \eqref{eq:z_m} are the same for  all the antennas, i.e., $a_{lm}=a_l$ for $m=1,\ldots,M$. Hence, the estimation quality is the same for all antennas and we need not to specify $M$ in the simulations related to the estimation performance. The simulation setup is the same as in Section \ref{se}. The pilot length is $\tau_{\text{p}}=K$ and the sequences are the columns of the discrete Fourier transform (DFT) matrix.

{\color{blue} \subsection{Training the Deep Neural Networks and Parameters}

The training data for both the neural networks in Fig.~\ref{fig:deep1} and \ref{fig:deep2} is generated by using the large-scale fading parameters according to the 3GPP Urban Microcell model in \cite{3gpp} with a
2\,GHz carrier frequency and 20\,MHz bandwidth. For each training sample, the users are dropped randomly in a cell of 250\,m$\times$250\,m. The large-scale fading coefficients, shadowing parameters, probability of LOS, and the Rician factors are simulated based on \cite[Table B.1.2.1-1, B.1.2.1-2, B.1.2.2.1-4]{3gpp} as in Section \ref{se}. Using the generated channels, the effective channels and the distortion variances are calculated using the derived results in Section \ref{effective}, \ref{se}, and \ref{general}. }
There are two hidden layers each with $30K$ neurons in the neural networks in Fig.~\ref{fig:deep1} and \ref{fig:deep2}. The mean squared error (MSE) is used as loss function. The first $2K$ inputs of the neural networks are scaled using the Standard Scaler and the others using the MinMax Scaler. {\color{blue} The scaling is needed for proper training and the motivation for these two types of scaling is as follows. The first $2K$ inputs can have both positive and negative values, hence Standard Scaler that removes the mean and normalize the input data such that it has unit variance is used for these inputs. On the other hand, the other $K$ inputs represent the square root of the channel gain over noise, which are always positive. Moreover, to prevent the large deviation between channel gains, these inputs are scaled between 0.1 and 0.9 using MinMax Scaler.} The outputs of the neural network in Fig.~\ref{fig:deep2} are also scaled using the MinMax Scaler, {\color{blue} which improves learning}. The Adam optimization algorithm is used {\color{blue} with learning rate 0.001 for training and the batch size and the maximum number of epochs are set as 1000 and 50, respectively.} The training and validation data lengths are $3\cdot10^{6}$ and $2\cdot10^{5}$, respectively. Some portion of the generated data corresponding to the outliers is not included in training which improves the learning. {\color{blue} The early stopping is applied by setting the patience parameter to 5, which is the number of epochs on which no improvement is seen in the validation loss.} 

\begin{figure}[t]
		\includegraphics[trim={3.8cm 0.35cm 6.7cm 3.4cm},clip,width=3.45in]{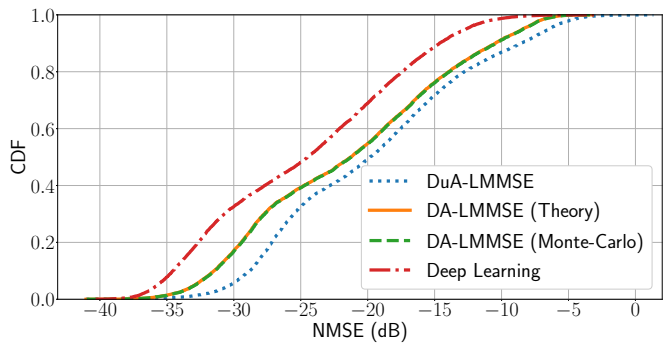}
	\caption{NMSE of effective channel estimation in dB for $K=10$ UEs and 3rd-order non-linear distortion with QPSK.}
	\label{fig:third_channel_10}
\end{figure}

{\color{blue} Based on the simulations carried out, we have empirically observed that increasing the number of neurons per layer results in better performance compared to increasing the depth of the neural networks. With the given parameters, significant performance improvement is obtained over the LMMSE-based methods. However, even better performance can be achieved by fine tuning the neural network and the training process, but this is left as future work.}
{\color{blue} \subsection{Computational Complexity} 
We note that the proposed deep neural networks are trained offline using data generated for a simulated cell with practical geometry. Since they are effectively trained to handle varying user SNRs and implicitly learning the SNR distribution of the considered propagation environment, the same networks can be used as long as the hardware impairment characteristics do not change. Hence, the main complexity of the proposed methods results from estimating the effective channels and distortion variances in testing stage. The computational complexity in testing a deep neural network is mainly determined by the number of layers and neurons per layer. For the considered neural networks, there are approximately $900K^2$ multiplications for each antenna per coherence block. 

For the distortion-unaware LMMSE, once the large-scale fading parameters are given, the complexity is determined by the simple scaling and addition in \eqref{eq:dist-un}. For the distortion-aware LMMSE, the coefficients of the matrices required for the effective channel estimation in \eqref{eq:dist-aw} are derived in closed-form in Section \ref{lmmse} for third-order non-linearities and they depend only on the large-scale fading parameters. For estimation of the small-scale effective channels, the computational complexity of distortion-aware LMMSE is determined by the matrix multiplications for each antenna element and user in \eqref{eq:dist-aw}. 

By only comparing the number of additions and multiplications, it is seen that LMMSE-based methods have less complexity. For a scenario with $K=10$ users and $M=100$ antennas, the average run time for distortion-aware LMMSE and deep neural network in Fig.~\ref{fig:deep1} is approximately 0.7 and 1.5 milliseconds without resorting to any parallel programming. Although distortion-aware LMMSE has lower computational time, deep learning does not add significant complexity, and as we will show in the next part, it provides significantly better performance improvement compared to the LMMSE-based benchmarks. Furthermore, there are no closed-form expressions for the matrices that are functions of large-scale fading parameters required for the distortion-aware LMMSE estimation of effective channels with non-linearities greater than order three and distortion variances. Hence, these matrices should be computed numerically. For deep learning, the proposed neural networks can be used similarly without an additional complexity since the only required large-scale fading parameters are the channel gains, that are given as inputs to the neural networks.  
}
{\color{blue} \subsection{Performance Comparison} }
Fig.~\ref{fig:third_channel_10} shows the normalized MSE (NMSE) of the effective channel estimates for $K=10$ users where the BS and UE hardware is modeled as a 3rd-order polynomial with QPSK modulation. There are 1000 different UE position setups where each point in Fig.~\ref{fig:third_channel_10} presents the average of 1000 channel realizations. DuA-LMMSE denotes the distortion-unaware LMMSE estimator in \eqref{eq:dist-un} and \cite{emil_nonideal}, and it acts as if the BS and UEs have ideal hardware. Hence, it has the worst performance among the considered ones. DA-LMMSE is the distortion-aware LMMSE which is derived in Section \ref{lmmse}. We compare it with the Monte-Carlo estimates and verify the correctness of the analytical expressions derived in Section \ref{lmmse}. As can be seen, the DA-LMMSE estimator outperforms DuA-LMMSE for each trial. However, the proposed deep-learning-based estimator provides substantially lower NMSE for almost each UE and setup. In fact, the median NMSE is improved by the proposed method by 3.2\,dB and 4.6\,dB compared to the DA-LMMSE and DuA-LMMSE estimators, respectively.

\begin{figure}[t]
	\includegraphics[trim={3.8cm 0.35cm 6.7cm 3.4cm},clip,width=3.45in]{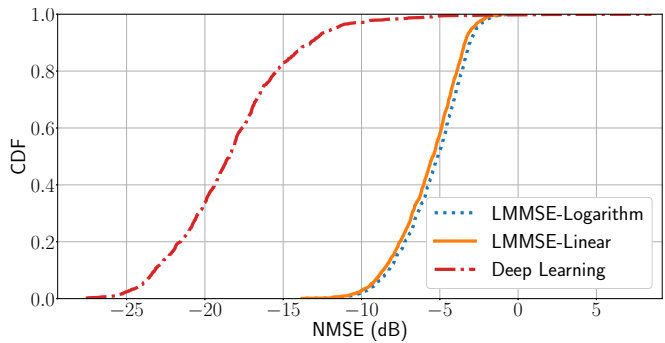}
	\caption{NMSE of distortion variance estimation in dB for $K=10$ UEs and 3rd-order non-linear distortion with QPSK.}
	\label{fig:third_distortion_10}
\end{figure}

In Fig.~\ref{fig:third_distortion_10}, we look at the estimation performance of the distortion variances, $[{\bf C}_{\mu\mu}]_{mm}$ for the same scenario. We first note that estimating the diagonal elements of the distortion correlation matrix using conventional correlation matrix estimation methods and effective channel estimates result in poor estimates. Hence, we restrict ourselves to compare the  estimation performance of the proposed deep learning-based method in Fig.~\ref{fig:deep2} with two schemes a) Monte-Carlo based LMMSE estimation of normalized distortion variance, $[{\bf C}_{\mu\mu}]_{mm}/\sigma^2$, and b) Monte-Carlo based LMMSE estimation of logarithm of distortion variance. The result of the method a) is converted to 1 if it is less than 1 using the knowledge $[{\bf C}_{\mu\mu}]_{mm}/\sigma^2\geq 1$ and it is denoted by LMMSE-Linear in Fig.~\ref{fig:third_distortion_10}. Similarly, the result of the method b) is converted to 0 if it is less than 0 using the same knowledge and it is denoted by LMMSE-Logarithm. Except for some very low probability outliers for very low SNR users, the proposed deep-learning based estimator outperforms these two LMMSE-based methods significantly (around 13\,dB improvement).

\begin{figure}[t]
	\includegraphics[trim={3.8cm 0.35cm 6.7cm 3.4cm},clip,width=3.45in]{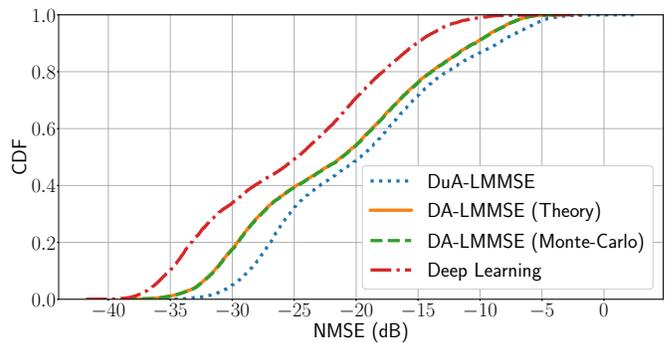}
	\caption{NMSE of effective channel estimation in dB for $K=10$ UEs and 3rd-order non-linear distortion with 16 QAM.}
	\label{fig:third_channel_10_16qam}
\end{figure}

We repeat the same experiment as in Fig.~\ref{fig:third_channel_10} for 16-QAM to show the robustness of the proposed approach to modulation differences. In fact, in addition to phase distortions, 16-QAM also suffers from amplitude distortions of constellation symbols at the UE transmitter. Fig.~\ref{fig:third_channel_10_16qam} shows the NMSE of the effective channel estimates for this scenario. Although Fig.~\ref{fig:third_channel_10_16qam} is very close to Fig.~\ref{fig:third_channel_10}, now the deep learning-based channel estimator provides more improvement, i.e., around 3.5\,dB and 5\,dB at the median point compared to DA-LMMSE and DuA-LMMSE showing the effectiveness of the proposed method.

\begin{figure}[t]
		\includegraphics[trim={3.8cm 0.35cm 6.7cm 3.4cm},clip,width=3.45in]{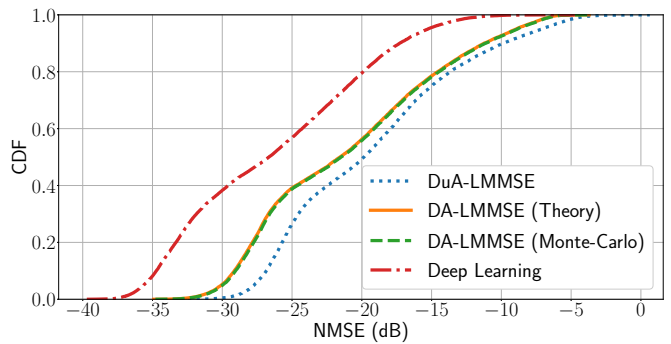}
	\caption{NMSE of effective channel estimation in dB for $K=20$ UEs and 3rd-order non-linear distortion with QPSK.}
	\label{fig:third_channel_20}
\end{figure}

\begin{figure}[t]
	\includegraphics[trim={3.8cm 0.35cm 6.7cm 3.4cm},clip,width=3.45in]{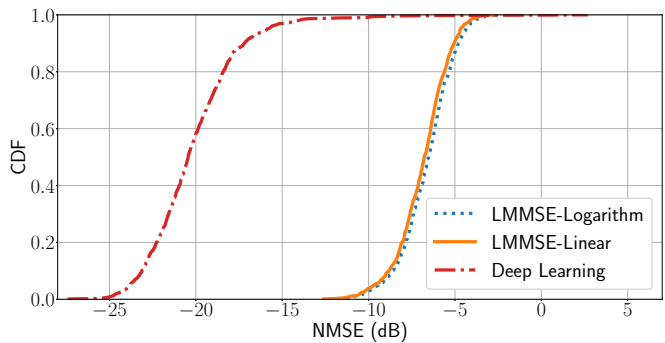}
	\caption{NMSE of distortion variance estimation in dB for $K=20$ UEs and 3rd-order non-linear distortion with QPSK.}
	\label{fig:third_distortion_20}
\end{figure}

Fig.~\ref{fig:third_channel_20} and Fig.~\ref{fig:third_distortion_20} denote the NMSE of effective channel and distortion variance estimates for $K=20$ UEs and QPSK modulation. Compared to $K=10$ UEs, we see that the performance gain between the conventional channel estimators and deep learning solution has increased and the median NMSE improvement is about 5 and 6.8\,dB compared to DA-LMMSE and DuA-LMMSE. We conclude that the proposed deep-learning-based estimator captures the structure of the hardware distortion which increases with the number of UEs, while the  LMMSE estimators fail to do so.

\begin{figure}[t]
	\includegraphics[trim={3.3cm 0.35cm 4.5cm 3.4cm},clip,width=3.45in]{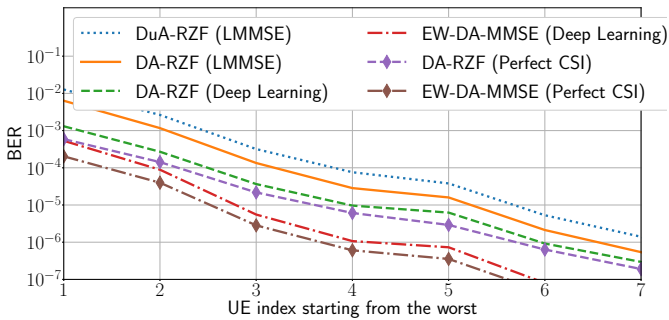}
	\caption{BER for $K=20$ UEs and 3rd-order non-linear distortion with QPSK.}
	\label{fig:third_ber_20}
\end{figure}

Fig.~\ref{fig:third_ber_20} shows the average uncoded bit error rate (BER) achieved by the DuA-RZF, DA-RZF, and EW-DA-MMSE receivers. The DuA-RZF receiver simply uses the distortion-unaware channel estimate to implement RZF. There are {\color{blue} three} DA-RZF receivers that are implemented with distortion-aware LMMSE, deep learning-based channel estimates {\color{blue} and perfect CSI}. The EW-DA-MMSE uses {\color{blue} either} the deep learning-based estimated effective channels and distortion variances {\color{blue} or the analytical results obtained with perfect CSI}. There are $M=100$ BS antennas with $K=20$ UEs. The average of 100 setups with random UE positions are plotted versus the UE index in ascending order of SNRs. 100 different channel realizations are considered per setup and 10,000 QPSK symbols are sent for each channel. As Fig.~\ref{fig:third_ber_20} shows, {\color{blue} the receivers that use perfect CSI always result smaller BER compared to the estimation-based schemes as expected. There is approximately 2-fold gap between perfect CSI-based receivers and deep learning-based estimation} and deep learning-based channel estimation improves the BER significantly compared to the LMMSE-based estimators. In fact, the BER reduction compared to DuA-RZF varies approximately between 4-fold and 10-fold wheres it is between 1.5-fold and 4-fold compared to DA-RZF with LMMSE-based estimate. Furthermore, using the diagonal elements of the distortion correlation matrix in EW-DA-MMSE improves the BER performance compared to DA-RZF with deep learning in a substantial manner. The gap increases with the UE index, hence SNR. In fact, there is more than a 10-fold BER reductions for the  6th and 7th UEs which shows the effectiveness of the proposed element-wise MMSE receiver.

\begin{figure}[t]
\includegraphics[trim={3.3cm 0.35cm 6.7cm 3.4cm},clip,width=3.45in]{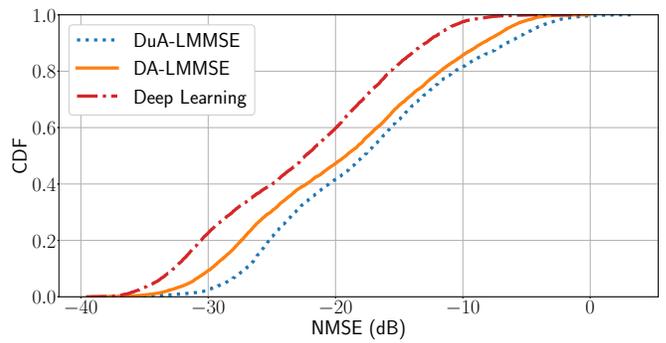}
	\caption{NMSE of effective channel estimation in dB for $K=10$ UEs and 7th-order non-linear distortion with QPSK.}
	\label{fig:seventh_channel_10}
\end{figure}

As a final simulation, we plot the NMSE of the channel estimates for  7th-order quasi-memoryless polynomial distortion in Fig.~\ref{fig:seventh_channel_10} in order to show the robustness of the proposed method. As it can be seen from this figure, the proposed deep-learning-based channel estimator provides a consistently better estimation quality by exploiting the hardware impairment structure.

\section{Conclusions}

We have analyzed the joint effect of non-linear distortions in the BS and UEs hardware on the estimation and detection in massive MIMO. The effective channels for any order of non-linearities and distortion correlation matrix for third-order non-linearities were analytically derived for the implementation of computationally efficient element-wise receivers for  uplink signal detection. SE of these distortion-aware receivers have been investigated and the statistics required for the  computation of LMMSE-based channel estimator is analytically derived for third-order non-linear distortions. Then, two new deep-learning-based channel  and distortion variance estimators were proposed. The neural networks were trained to utilize the hardware distortion characteristics to achieve better estimation quality than with the conventional Bayesian LMMSE estimators used in the massive MIMO literature, which treat the distortion as an independent colored noise and only utilizes its first- and second-order statistics. We have shown that the same neural networks trained offline can be utilized to provide significantly better estimates in practical Rician fading channel setups with varying SNRs. Moreover, the proposed deep-learning based estimators only require the channel gain information and do not require the separate estimation of LOS components which brings big practical advantage. 

In summary, we have shown how the data-driven deep-learning approach can be combined with expert-knowledge from the wireless communication field to exploit the structure of transceiver hardware and thereby outperform previous suboptimal model-based designs.

\appendices
\section{Proof of Lemma 1}
Let us consider the first case in \eqref{eq:symbol-gaussian}, where we have $l_1=l_2=l_3=k$ which results in 
\begin{align}\label{eq:app_a1}
\mathbb{E}\{\upsilon_{l_1}\upsilon_{l_2}^{*}\upsilon_{l_3}\varsigma_k^{*}\}=&\mathbb{E}\{|\upsilon_{k}|^2\upsilon_{k}\varsigma_k^{*}\}\nonumber\\
=&\mathbb{E}\bigg\{\big|\tilde{b}_{0}\varsigma_k+\tilde{b}_{1}|\varsigma_k|^2\varsigma_k\big|^2(\tilde{b}_{0}|\varsigma_k|^2+\tilde{b}_{1}|\varsigma_k|^4)\bigg\} \nonumber \\
\stackrel{(a)}{=}&\zeta_{10}B_{1,1,1}+2\zeta_8B_{1,1,0}+\zeta_8B_{1,0,1}\nonumber\\
&+2\zeta_6B_{0,0,1}+\zeta_6B_{0,1,0}+\zeta_4B_{0,0,0},
\end{align}
where we used the definitions in \eqref{eq:moments1} and \eqref{eq:B2} in (a). The second and third cases in \eqref{eq:symbol-gaussian} can be proved similarly by using independence of data signals for different users. The last case follows directly from that the data signals satisfy the $90^{\circ}$ circular shift symmetry.

\section{Proof of Lemma 2}
Let us define ${\bf R}\triangleq\mathbb{E}\{\bm{\upsilon}\bm{\upsilon}^H{\bf A}\bm{\upsilon}\bm{\upsilon}^H\}$. The $(i,j)$th element of ${\bf R}$ for $i\neq j$ is given by
\begin{align} \label{eq:Rij}
[{\bf R}]_{ij}=\sum_{p=1}^K\sum_{r=1}^KA_{pr}\mathbb{E}\{\upsilon_i\upsilon_r\upsilon_p^{*}\upsilon_j^{*}\}=\chi_2^2A_{ij}, \ \ i\neq j,
\end{align}
where $A_{pr}=[{\bf A}]_{pr}$ is the $(p,r)$th element of the matrix ${\bf A}$ and we used the $90^{\circ}$ circular shift symmetry together with $i\neq j$. The diagonal elements of ${\bf R}$ is given by
\begin{align} \label{eq:Rii}
[{\bf R}]_{ii}=\sum_{p=1}^K\sum_{r=1}^KA_{pr}\mathbb{E}\{\upsilon_i\upsilon_r\upsilon_p^{*}\upsilon_i^{*}\}=\chi_4A_{ii}+\sum_{p\neq i}^K\chi_2^2A_{pp}.
\end{align}
Using these results, ${\bf R}$ is given as in \eqref{eq:lemma2-2}.

Let us consider the second claim of Lemma 2. If we define ${\bf S}\triangleq\mathbb{E}\{\bm{\upsilon}\bm{\upsilon}^H{\bf A}\bm{\upsilon}\bm{\upsilon}^H{\bf  B}\bm{\upsilon}\bm{\upsilon}^H\}$, the $(i,j)$th element of ${\bf S}$ for $i\neq j$ is given by
\begin{align} \label{eq:Sij}
[{\bf S}]_{ij}=&\sum_{p=1}^K\sum_{r=1}^K\sum_{l=1}^K\sum_{n=1}^KA_{pr}B_{ln}\mathbb{E}\{\upsilon_i\upsilon_r\upsilon_n\upsilon_p^{*}\upsilon_l^{*}\upsilon_j^{*}\}\nonumber\\
=&\chi_4\chi_2\bigg(A_{ii}B_{ij}+B_{ii}A_{ij}+A_{jj}B_{ij}+B_{jj}A_{ij}\bigg) \nonumber\\
&+\chi_2^3\bigg(A_{ij}\sum_{\substack{n\neq i\\ n \neq j}}^KB_{nn}+B_{ij}\sum_{\substack{n\neq i\\ n \neq j}}^KA_{nn}\nonumber\\
&+\sum_{\substack{n\neq i\\ n \neq j}}^KA_{in}B_{nj}+\sum_{\substack{n\neq i\\ n \neq j}}^KB_{in}A_{nj}\bigg), \ \ i\neq j.
\end{align}
The diagonal elements of ${\bf S}$ are given by
\begin{align} \label{eq:Sii}
[{\bf S}]_{ii}=&\sum_{p=1}^K\sum_{r=1}^K\sum_{l=1}^K\sum_{n=1}^KA_{pr}B_{ln}\mathbb{E}\{\upsilon_i\upsilon_r\upsilon_n\upsilon_p^{*}\upsilon_l^{*}\upsilon_i^{*}\}\nonumber\\
=&\chi_6A_{ii}B_{ii}+\chi_4\chi_2\sum_{n\neq i} A_{nn}B_{nn} \nonumber \\
&+\chi_4\chi_2\bigg( A_{ii}\sum_{\substack{n\neq i}}^KB_{nn}+B_{ii}\sum_{\substack{n\neq i}}^KA_{nn}\nonumber\\
&+\sum_{\substack{n\neq i}}^KA_{in}B_{ni}+\sum_{\substack{n\neq i}}^KB_{in}A_{ni}\bigg)\nonumber\\
&+\chi_2^3\bigg(\sum_{p\neq i}^K\sum_{\substack{n\neq p\\ n \neq i}}^K\big(A_{pp}B_{nn}+A_{pn}B_{np}\big)\bigg).
\end{align}
After arranging the terms in \eqref{eq:Sij} and \eqref{eq:Sii}, the result in Lemma 2 can be obtained as in \eqref{eq:lemma2-4}.

\end{document}